\documentclass[12pt]{iopart}
\usepackage{times}
\usepackage{epsfig}
\usepackage{graphicx}
\usepackage{dcolumn}
\usepackage{natbib}

\def\dj{\hbox{d\kern-0,347em \vrule width0,3em height1,252ex
depth-1,21ex \kern0,051em}}
 
\begin{document}

\title{Percolation and localization in the random fuse model}

\author{Phani Kumar V.V. Nukala\dag, Sr{\dj}an \v{S}imunovi\'{c}\dag, Stefano Zapperi\ddag}
\address{\dag Computer Science and Mathematics Division, 
Oak Ridge National Laboratory, Oak Ridge, TN 37831-6359, USA}
\address{\ddag INFM UdR Roma 1 and SMC, Dipartimento di Fisica,
Universit\`a "La Sapienza", P.le A. Moro 2, 00185 Roma, Italy}
 
\begin{abstract}
We analyze damage nucleation and localization in the random fuse model
with strong disorder using numerical simulations. In the initial stages
of the fracture process, damage evolves in an uncorrelated manner, 
resembling percolation. Subsequently, as the damage starts to accumulate, 
current enhancement at the tips of the microcracks leads eventually to 
catastrophic failure. We study this behavior quantifying
the deviations from percolation and discussing alternative scaling 
laws for damage. The analysis of damage profiles confirms that 
localization occurs abruptly starting from an uniform damage landscape.
Finally, we show that the cumulative damage distribution follows
the normal distribution, suggesting that damage is uncorrelated on
large length scales. 
\end{abstract}

%\maketitle

\section{Introduction}

Understanding the scaling properties of fracture in disordered media
represents an intriguing theoretical problem with important implications
for practical applications \cite{breakdown}. Experiments have shown that in several materials
under different loading conditions, the fracture surface is 
rough and can be described by self-affine scaling \cite{man}
with universal exponents \cite{bouch}. Scaling is also observed in
acoustic emission experiments, where the distribution of pulses
decays as a power law over several decades.
Experimental observations have been reported for several 
materials such as wood \cite{ciliberto}, cellular glass
\cite{strauven}, concrete \cite{ae} and paper \cite{paper},
but universality does not seem to hold.The experimental observation of power law behavior suggests 
an interpretation in terms of critical phenomena and scaling theories,
but a complete theoretical explanation has not been found.
The statistical properties of fracture in disordered media
are often studied  with lattice models, describing the medium as
a discrete set of elastic bonds with random failure thresholds
\cite{breakdown}. These 
numerical simulations are used in estimating the roughness of 
the fracture surface, which is found to be self-affine
\cite{hansen91b,bat-98,rai-98,sep-00}, and 
the power law distribution of avalanche precursors 
\cite{hansen,zrvs,zvs,gcalda,alava}.
While the results agree qualitatively with experiments a 
quantitative comparison is not always satisfactory.

Apart from the comparison with experiments, an important theoretical issue
is to understand the origin of the scaling behavior observed in 
the numerical simulations of lattice model. A very well studied model
is the random fuse model (RFM), where a lattice of fuses with
random threshold are subject to an increasing voltage 
\cite{deArcangelis85,breakdown,zrvs,kahng88,delaplace,deArcangelis89,nukalajpamg}.
A resistor network represents a scalar analog of an elastic medium
and is thus relatively simple to analyze, while retaining some important
characteristic features of the problem. 

Simulations of the RFM show that the type of behavior at
macroscopic fracture is significantly influenced by the amount of
disorder \cite{kahng88}.  When the disorder is narrowly distributed,
materials breakdown without significant precursors. As the disorder
increases, substantial damage is accumulated prior to failure and
the dynamics resembles percolation \cite{sornette}. 
Indeed, in the limit of infinite disorder, the damage
accumulation process can exactly be mapped onto a percolation problem
\cite{guyon88}. It is still debated, however, if percolation
is applicable to the case of strong but  non-infinite disorder.
Recently, some evidence has been provided in this direction \cite{hansen003}
suggesting that the critical exponent, $\nu$, of the correlation length in the RFM with 
strong disorder is same as that of uncorrelated percolation (i.e. $\nu = 4/3$), and that 
the fuse model is in the same universality class of percolation. 
Close to failure damage would then localize 
and the resulting crack roughness would ensue from a gradient percolation
mechanism \cite{hansen003}. 

In the present work, we propose a different interpretation of 
damage localization in the RFM: while in the initial stages damage accumulates 
as in a percolation process, the final crack nucleates abruptly 
due to current enhancement, yielding the observed localization profiles. 
As a consequence of this, the damage localization length follows closely the scaling
of crack width. In addition, we test percolation scaling by simulating the 
RFM  with triangular and diamond (square lattice
with 45 degrees inclined bonds to the bus bars) lattice topologies
\cite{deArcangelis85,breakdown} and different disorder distributions
(uniform and power law). The numerical results allows us to exclude that
$\nu = 4/3$, but could still be compatible with another value of $\nu$,
although devations from scaling are observed for large sizes.

Finally, we show that the cumulative damage probability distribution 
at failure can be collapsed for different lattice sizes and follows the normal
distribution. This suggests that damage is the sum of uncorrelated
variables and thus no long-range correlation are present as would be
expected in critical phenomena. This fact, together with the abrupt
localization occurring right at failure, suggests the validity
of the first-order transition hypothesis discussed in Ref.~\cite{zrvs}.

The paper is organized as follows: in section II we define the
model and in section III we discuss the role of disorder. The
percolation hypothesis is tested in section IV, while section
V is devoted to damage localization. The scaling of
the number of broken bonds is discussed in section VI. Section
VII presents the analysis of the failure probability distribution
and section VIII summarizes briefly our conclusions.

\section{Model}

In the random thresholds fuse model, the lattice is initially fully intact
with bonds having the same conductance, but the bond breaking
thresholds, $t$, are randomly distributed based on a thresholds
probability distribution, $p(t)$.  The burning of a fuse occurs
irreversibly, whenever the electrical current in the fuse exceeds the
breaking threshold current value, $t$, of the fuse. Periodic boundary
conditions are imposed in the horizontal direction to simulate an
infinite system and a constant voltage difference, $V$, is applied
between the top and the bottom of lattice system bus bars.

Numerically, a unit voltage difference, $V = 1$, is set between the
bus bars and the Kirchhoff equations are solved to determine the
current flowing in each of the fuses. Subsequently, for each fuse $j$,
the ratio between the current $i_j$ and the breaking threshold $t_j$
is evaluated, and the bond $j_c$ having the largest value,
$\mbox{max}_j \frac{i_j}{t_j}$, is irreversibly removed (burnt).  The
current is redistributed instantaneously after a fuse is burnt
implying that the current relaxation in the lattice system is much
faster than the breaking of a fuse.  Each time a fuse is burnt, it is
necessary to re-calculate the current redistribution in the lattice to
determine the subsequent breaking of a bond.  The process of breaking
of a bond, one at a time, is repeated until the lattice system falls
apart. In this work, we consider a uniform probability distribution, 
which is constant between 0 and 1, and a power law distribution
$p(t) \propto t^{-1+\beta}$, $0 \le t \le 1$ with
$\beta = \frac{1}{20}$.

Numerical simulation of fracture using large fuse networks is often
hampered due to the high computational cost associated with solving a
new large set of linear equations every time a new lattice bond is
broken.  The authors have developed rank-1 sparse Cholesky 
factorization updating algorithm
for simulating fracture using discrete lattice systems
\cite{nukalajpamg}.  In comparison with the Fourier accelerated
iterative schemes used for modeling lattice breakdown
\cite{batrouni}, this algorithm significantly reduced the
computational time required for solving large lattice systems. Using
this numerical algorithm, we were able to investigate damage
evolution in larger lattice systems (e.g., $L = 1024$), which to the
authors knowledge, is so far the largest lattice system used in studying
damage evolution using initially fully intact discrete lattice systems. 
However, in this paper, we consider results up to $L = 512$ due to insufficient number 
of available sample configurations for $L = 1024$, which will be considered 
in a future publication. 

Using the algorithm presented in \cite{nukalajpamg}, we have performed
numerical simulations on two-dimensional triangular and diamond
(square lattice inclined at 45 degrees between the bus bars) lattice
networks.  For many lattice system sizes,
the number of sample configurations, $N_{config}$, used are
excessively large to reduce the statistical error in the numerical
results (see Tables 1 and 2).  Each numerical simulation was performed on a single
processor of {\it Eagle} (184 nodes with four 375 MHz Power3-II
processors) supercomputer at the Oak Ridge National
Laboratory. The statistically independent $N_{config}$ number
of configurations were simulated simultaneously on number of
processors available for computation.

\section{The role of disorder}

The disorder distribution has clearly an important effect on the
fracture behavior and a classification has been proposed in terms
of a scale-invariant spectrum \cite{hansen91,hansen001}, which
for the uniform thresholds distribution $(0 \le t \le 1)$ in terms of intensive
variables, $\alpha_t$ and $f_t (\alpha_t)$, is given by
\cite{hansen91,hansen001}
\begin{equation}
f_t (\alpha_t) = 2 - \alpha_t ~~~~ \mbox{for} ~~ 0 \le \alpha_t \le
2 \label{unif_spec}
\end{equation}
where 
\begin{eqnarray}
\alpha_t & = & \frac{\mbox{log} ~t}{\mbox{log} ~L} \label{alphat} \\
f_t (\alpha_t) & = & \frac{\mbox{log} ~L^2 t p(t)}{\mbox{log} ~L}
\label{ftalphat}
\end{eqnarray}
Hence, for the uniform distribution between 0 and 1, the two control
parameters $\phi_0$, and $\phi_{\infty}$ that characterize the
thresholds distribution $p(t)$ close to zero and infinity,
respectively, are given by $\phi_0 = 1$, and $\phi_{\infty} =
\infty$ \cite{hansen91,hansen001}, where $\phi_{0/\infty}$ are defined by 
\begin{equation}
\phi_{0/\infty} = \lim_{t\to0/\infty}{\left( \frac{\log(t p(t))}{\log (t)}\right)}
\end{equation} 

For the power law thresholds distributions as the one we are using, the
scale-invariant thresholds spectrum is given by
\begin{equation}
f_t (\alpha_t) = 2 - \beta \alpha_t ~~~~ \mbox{where} ~~ 0 \le
\alpha_t \le \frac{2}{\beta} \label{power_spec}
\end{equation}
and the two control parameters are $\phi_0 = \beta$ and $\phi_{\infty}
= \infty$ \cite{hansen91,hansen001}. According to Ref.~ \cite{hansen91,hansen001}, 
based on the values of
these two control parameters, $\phi_0$ and $\phi_{\infty}$, both the
uniform thresholds distribution and power-law thresholds
distributions (as long as $\beta<2$ \cite{hansen001}) belong to the same scaling
regime characterized by diffusive damage and localization (see regime
B of Figure 18 in reference \cite{hansen001}). 
According to this analysis,  if the exponent $\nu$ were to 
be universal and is in the same universality
class as that of uncorrelated percolation as conjectured in
\cite{hansen003}, then we expect to find
$\nu=4/3$ for uniform and power law threshold distributions. 

\section{Damage and percolation}

In the case of strong disorder, in the initial stages of damage evolution, 
bond breaking events occur in an uncorrelated manner and thus resemble percolation. 
As the damage starts to accumulate, some degree of correlation
can be expected because of the current enhancement present at the crack tips.
A natural question to ask concerns the relevance of these correlations
as failure is approached. If correlations are irrelevant one should observe percolation
scaling up to failure, as in the case of infinite disorder. On the other hand, in the 
weak disorder case, the current enhancement at the crack tips is so strong that 
a spanning crack is nucleated soon after a few bonds (or even a single bond) are broken \cite{kahng88}. 
The interesting situation corresponds to the diffuse damage and localization regime,
where a substantial amount of damage is accumulated prior to failure.
In this regime one should test if damage follows percolation scaling
up to failure. 

Hansen and Schmittbuhl \cite{hansen003} have considered broad  
threshold distributions $p(t) \propto t^{-1+\beta}$, $0 \le t \le 1$ 
with two different $\beta = \frac{1}{10}$ and $\beta =
\frac{1}{3}$ values, to see if percolation scaling was observed for $\beta>0$.
Based on the similarities with percolation, they \cite{hansen003} 
suggested the following finite size scaling law for the fraction of
broken bonds, given by
\begin{eqnarray}
p_{f} - p_c & \sim &  L^{-\frac{1}{\nu}} \label{pf}
\end{eqnarray}
In the Eq. (\ref{pf}), $p_{f}$ and $p_c$ represent the fracture
thresholds in a lattice system size of $L$ and infinity, respectively.
As the system size $L \rightarrow \infty$, the
broken bonds at failure $p_{f} \rightarrow p_c$.
The correlation critical exponent $\nu$ was found in Ref.~\cite{hansen003}
to be consistent with the percolation value $\nu=4/3$.
An additional test is provided by the damage standard deviation at
failure $\Delta_f$ \cite{ramstad} which should scale as 
\begin{equation}
\Delta_f \sim L^{-\frac{1}{\nu}}
\label{eq:delta}
\end{equation}

\begin{figure}[hbtp]
\centerline{\includegraphics[width=12cm]{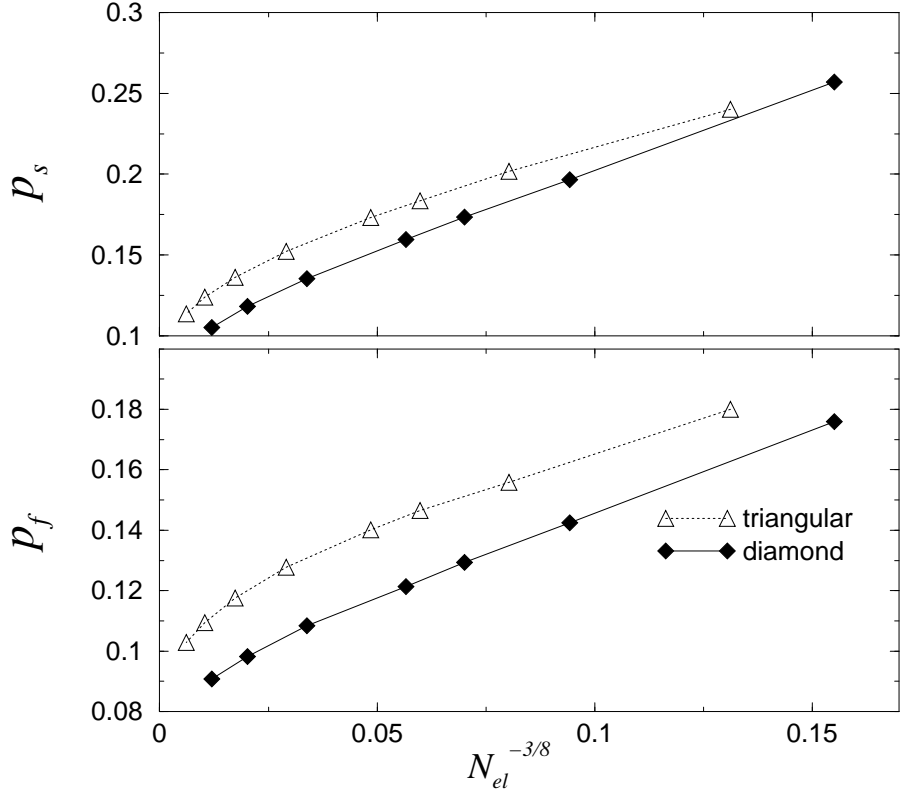}}
\caption{The 50\% survival probability $p_s$ (top) and the mean 
fraction of broken bonds $p_{f}$ (bottom) plotted as a function
of $N_{el}^{-3/8}$ for the uniform threshold distribution.
 If percolation scaling is obeyed, the data should
follow a straight line. A net curvature is instead observed in
all the data for large lattice sizes.}
\label{fig:1}
\end{figure}

Here we test these scaling laws for a wider finite size
range than Ref.~\cite{hansen003}, wherein simulations with sizes up to
$L=60$ are reported.  The fraction of broken bonds for each of the
lattice system sizes is obtained by dividing the number of broken
bonds with the total number of bonds, $N_{el}$, present in the fully
intact lattice system. For triangular lattice topology, $N_{el} =
(3L+1) (L+1)$, and for diamond lattice topology, $N_{el} = 2L (L+1)$.
The lattice system sizes considered in this work are $L = \{8, 16, 24,
32, 64, 128, 256, 512\}$.  However, since corrections to the scaling
laws are strongest for small lattice systems, in the following, we use
lattice sizes $L \ge 16$ for obtaining the scaling exponents.  Table 1
presents mean and standard deviations in the broken bond density
(fraction of broken bonds) at the peak load and at failure for various
lattice system sizes in both the triangular and diamond lattice
systems for uniform thresholds distribution.  In order to compare
diamond and triangular lattice topologies, we find it more convenient
to use $N_{el}$ rather than $L$ as a finite size parameter. This is
because the two lattices have a different dependence of the real
lattice size (i.e., $N_{el}$) on the linear size $L$.  We plot in
Fig.~\ref{fig:1}, the mean fraction of broken bonds at failure $p_{f}$
as a function of $N_{el}^{-3/8}$ for diamond and triangular lattices,
which in principle should obey Eq.~\ref{pf} as well. While to accept
the percolation hypothesis one should observe a linear regime, a net
curvature is apparent in the data especially for sizes $L>100$.  A
similar curvature is found in the 50\% survival probability
$p_s$. 

\begin{figure}[hbtp]
\centerline{\includegraphics[width=12cm]{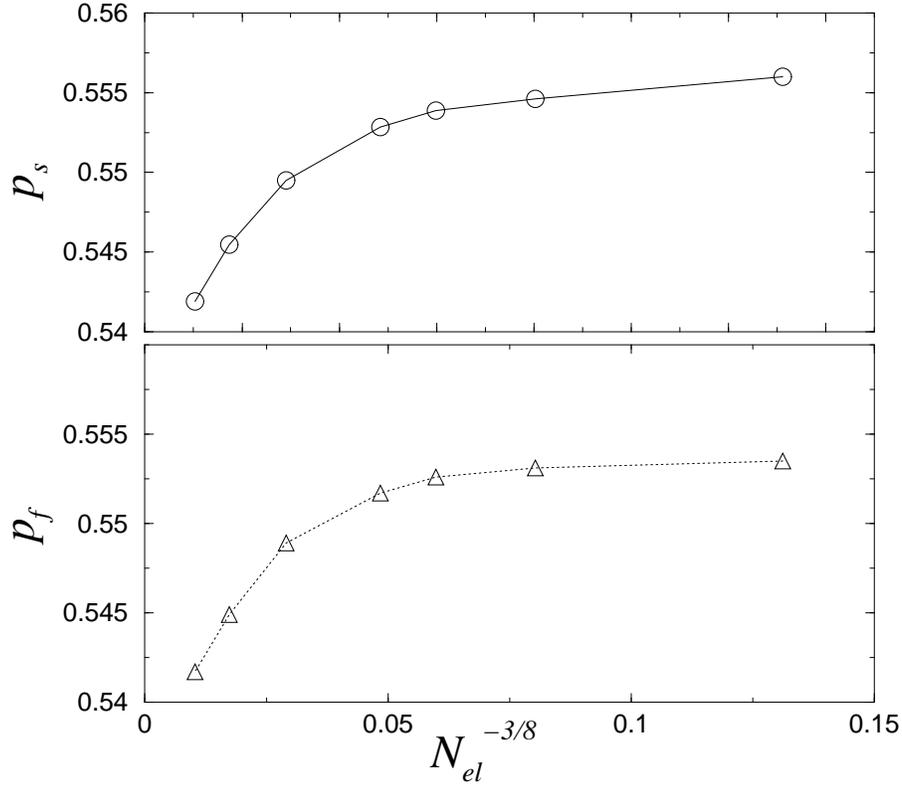}}
\caption{The 50\% survival probability $p_s$  and the mean 
fraction of broken bonds $p_{f}$  plotted as a function
of $N_{el}^{-3/8}$ for the power law threshold distribution. 
Percolation scaling is not obeyed.}
\label{fig:1b}
\end{figure}

We repeated the same analysis using the power law
threshold distribution with $\beta=1/20$ in triangular lattices of sizes 
$L=8,16,24,32,64,128,256,512$. Table 2 
presents mean and standard deviations in the broken bond density 
(fraction of broken bonds) at the peak load and at failure for 
power law thresholds distribution. The result is reported 
in Fig.~\ref{fig:1b}, and is once again in contrast with percolation scaling (i.e. $\nu\ne 4/3$)

\begin{figure}[hbtp]
\centerline{\includegraphics[width=12cm]{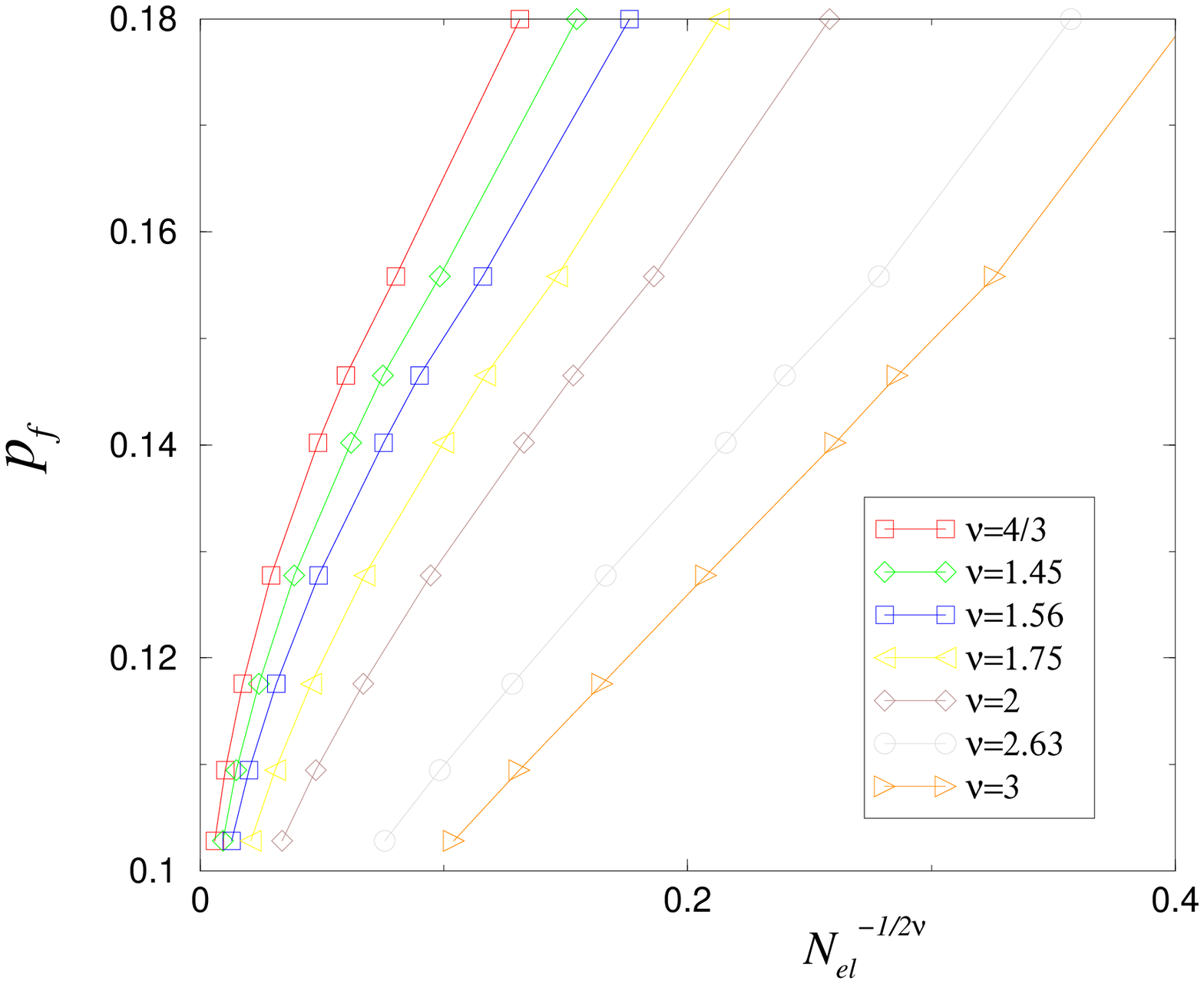}}
\centerline{\includegraphics[width=12cm]{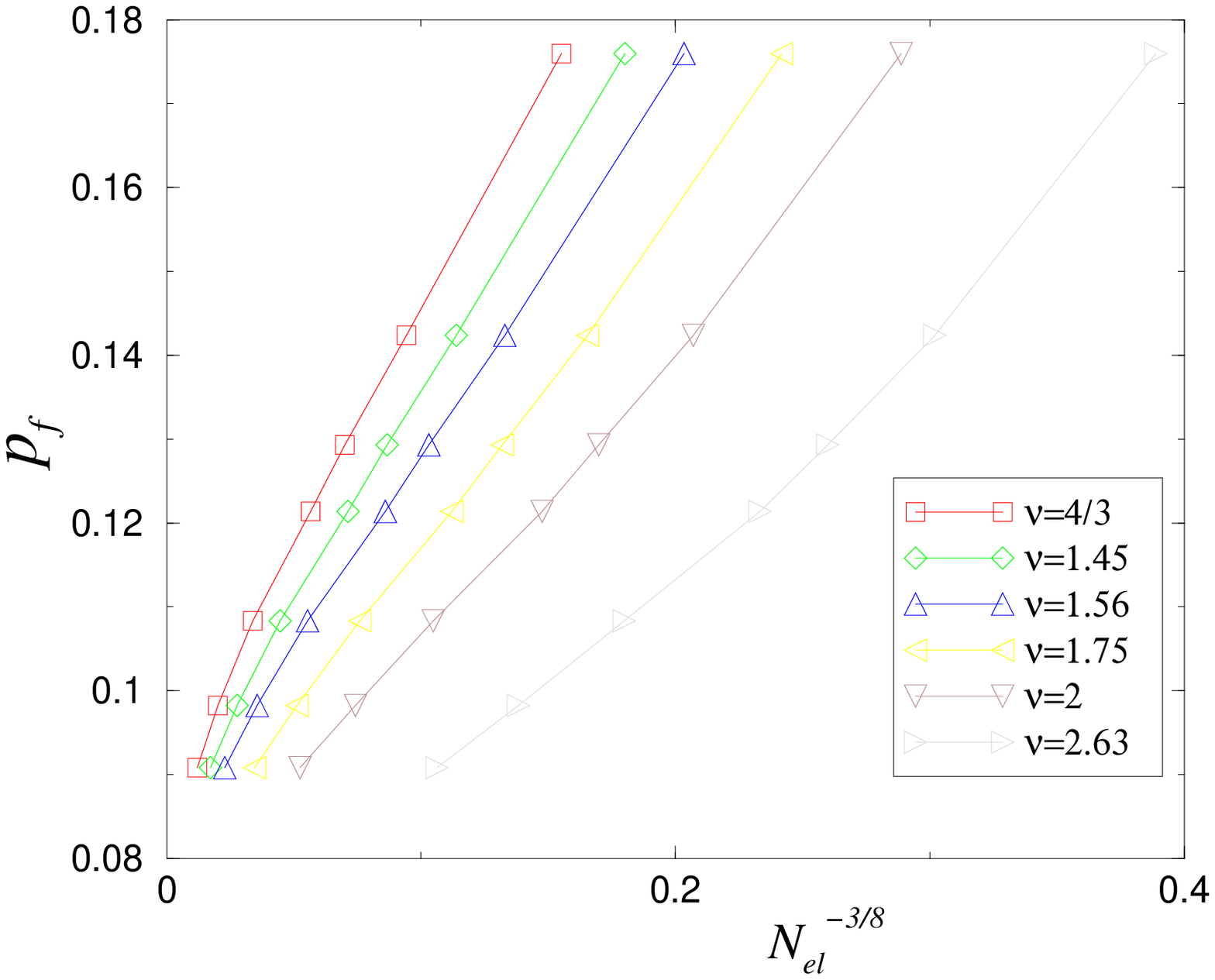}}
\caption{The mean 
fraction of broken bonds $p_{f}$ plotted as a function
of $N_{el}^{-1/2\nu}$ for the triangular (top) and diamond (bottom)
lattice with a uniform threshold distribution. A straight line is 
observed for different values of $\nu$ in the two case.}
\label{fig:1c}
\end{figure}

Thus we can conclude that in the random fuse model with a
uniform and power law distributions $\nu$ is not equal to $4/3$.  
On the basis of the results, however, we can not exclude the possibility that
Eq. (\ref{pf}) is valid with a different value of $\nu$. This would
correspond to some sort of correlated percolation, which, being a
second order critical phenomenon, is expected to be universal with
respect to the lattice structure. A way to test this idea is to
plot the data as in Figs.~\ref{fig:1}-\ref{fig:1b} using a different
value of $\nu$. In Figs.~\ref{fig:1c}a and ~\ref{fig:1c}b we report similar plots
for the triangular and diamond lattice topologies with uniform thresholds distribution, where it can be seen
that a straight line is obtained only for large values of $\nu$. In addition, the $\nu$ values 
for which a straight line is obtained in Figs.~\ref{fig:1c}a and ~\ref{fig:1c}b are quite 
different from one another. For example, a direct
fit of the data for $L \ge 16$ yields
\begin{eqnarray}
p_{f} - 0.0816 & = & 0.42 ~ N_{el}^{-0.19},\label{pfcfit} 
\end{eqnarray}
for triangular lattices, and 
\begin{eqnarray}
p_{f} - 0.0751 & = & 0.57 ~ N_{el}^{-0.25}.\label{pfcfitd}
\end{eqnarray}
for diamond lattices. 
Hence, Eqs. (\ref{pfcfit}) and (\ref{pfcfitd}) for triangular and
diamond lattice topologies, estimate the scaling exponent $\nu$ to be
equal to $2.63$ and $2.0$, respectively.

\begin{figure}[hbtp]
\centerline{\includegraphics[width=12cm]{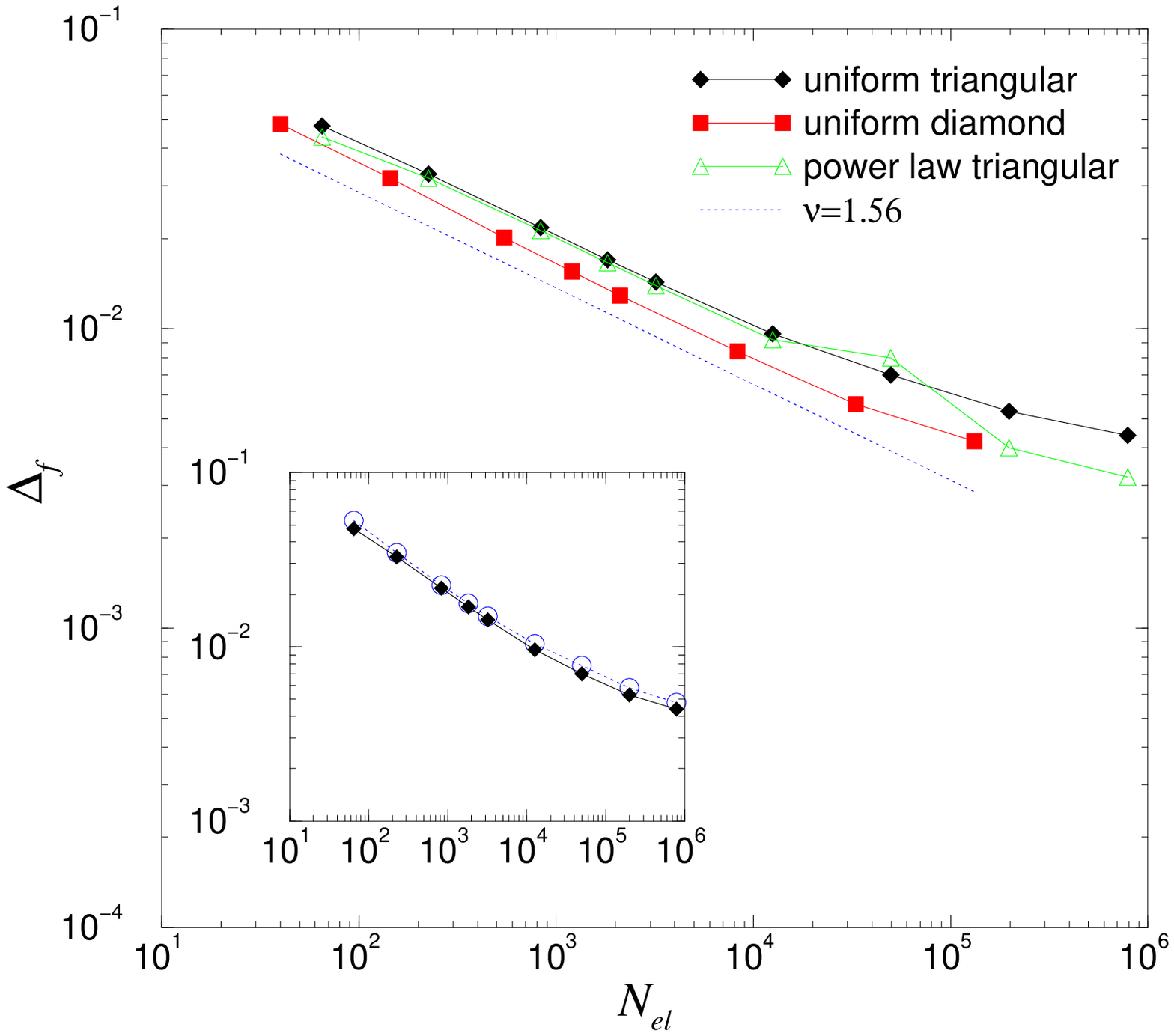}}
\caption{The standard deviation $\Delta_{f}$ of the
fraction of broken bonds at failure plotted as a function
of $N_{el}$ for different lattices and disorder distributions.
The curves are initially described by a power law with exponent
$-0.32$ corresponding to $\nu=1.56$, but flatten for
$L>100$. In the inset, we compare $\Delta_f$ with the same
quantity computed at peak load in the case of a triangular
lattice with uniform disorder.}
\label{fig:delta}
\end{figure}

An additional test is provided by plotting the standard deviation
of bonds at failure, which in case of percolation
should follow Eq.~\ref{eq:delta}. For small lattice sizes the data follow
a power law with an exponent $\nu\simeq 1.56$, but deviations occur at large
sizes. In the framework proposed in Ref.~\cite{hansen003}
one could interpret the data in Fig.~\ref{fig:delta} as an indication of
an initial correlated percolation process with $\nu=1.56$, which then crosses over
to a localized state where scaling is obscured by the presence of a 
damage concentration profile. We will show in the next section, however,
that there is apparently no localization at peak load (i.e. the maximum current
before catastrophic failure). On the other hand, the damage  
standard deviation at peak load follows closely the behavior at
failure as shown in the inset of Fig.~\ref{fig:delta}. Thus the
damage profile does not appear to be responsible from the deviation
from scaling. A different explanation would be that
there is no scaling just because the lattice fails abruptly 
far from a (correlated) percolation  critical point.

\section{Damage localization}

As discussed in the previous section, damage is not described by 
percolation critical scaling up to the largest sizes. Fracture
is abrupt and damage localizes. Here we clarify when and how localization
takes place. In particular, we will consider the damage accumulated up
to the peak load (i.e. the highest current that the lattice can bear
without breaking) and after the peak load up to failure.

\begin{figure}[hbtp]
\centerline{\psfig{file=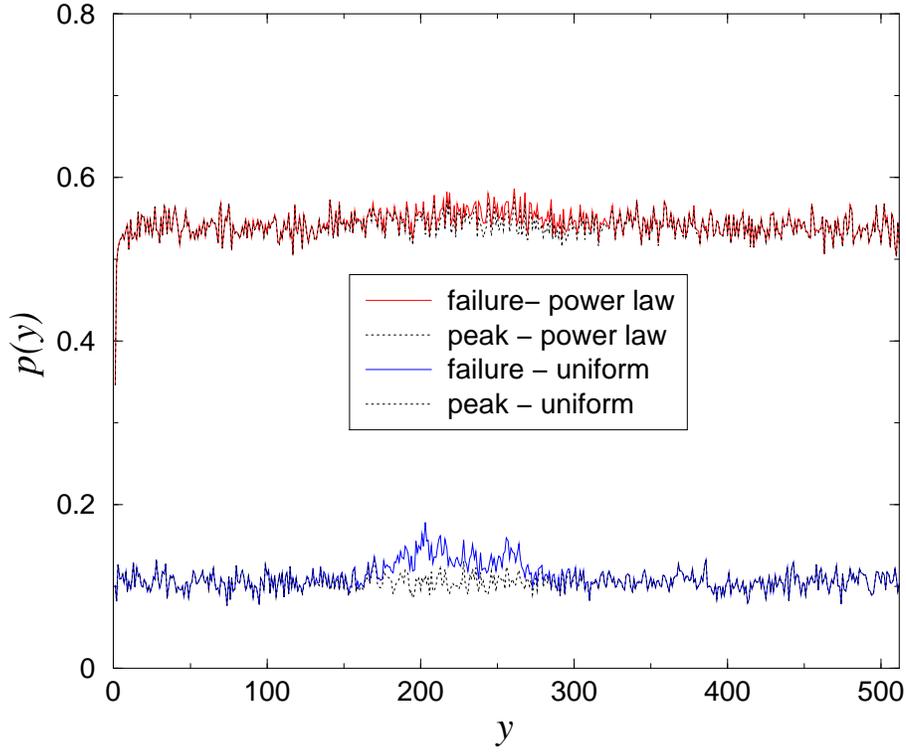,width=12cm,clip=!}}
\caption{Normalized damage profile $p(y) = \frac{n_b(y)}{(3 L + 1)}$
at failure and at peak load in a typical triangular lattice simulation
of size $L = 512$ for the cases of uniform and power law
disorders. $n_b(y)$ denotes the number of broken bonds in the $y^{th}$
section.}
\label{fig:loc1}
\end{figure}

In Fig.~\ref{fig:loc1} we display the damage profile $p(y)$ at failure
and at peak load in a single simulation, for the cases of uniform and
power law disorders. For uniformly distributed thresholds,
localization appears clearly at failure, while at peak load the damage
spreads homogeneously though the lattice. For power law disorder, it is
difficult to assess from these curves the extent of localization.
To obtain a quantitative description of the localization process
it is necessary to average over different realizations.

Averaging the profiles is a delicate task since localization does not
necessarily take place in the center, but can in principle occur
anywhere along the length of the lattice.  Thus one can not
perform a simple average because this would yield a flat average profile
irrespective of the individual profile shapes in a single realization.  
The authors of Ref.~\cite{mikko04}
proposed to first shift the profiles so that they are centered around
the maximum and then average. This method emphasizes the noise too
much, yielding a spurious cusp in the center.  Another possibility is to shift
instead by the center of mass of the damage or, to avoid any effects due
to shifting, one can use the Fourier method.

\begin{figure}[hbtp]
\centerline{\psfig{file=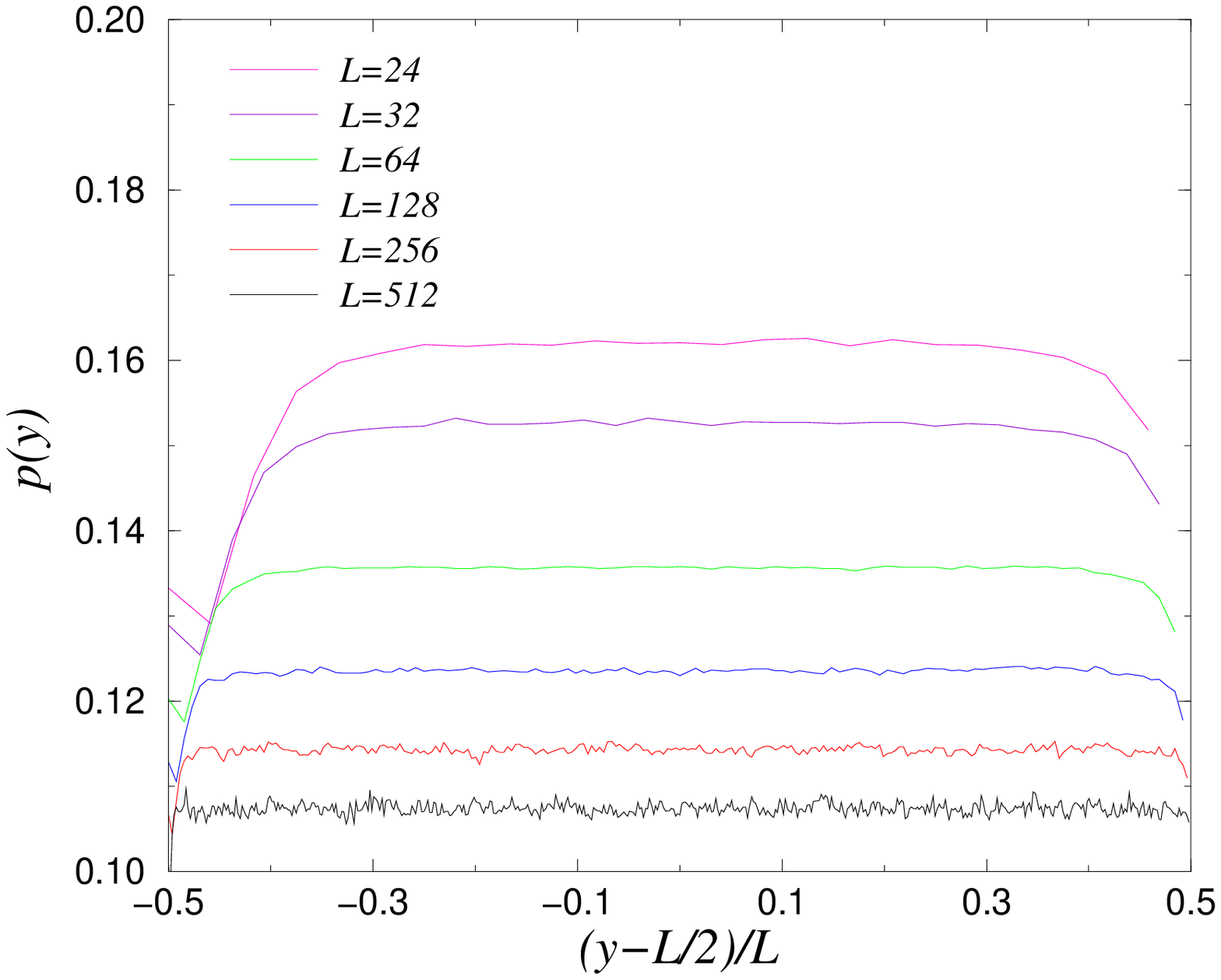,width=12cm,clip=!}}
\caption{Average damage profiles at peak load obtained by first centering the data
around the center of mass of the damage and then averging over different samples.
The data correspond to unifrormly distributed disorder.}
\label{fig:loc2}
\end{figure}

\begin{figure}[hbtp]
\centerline{\psfig{file=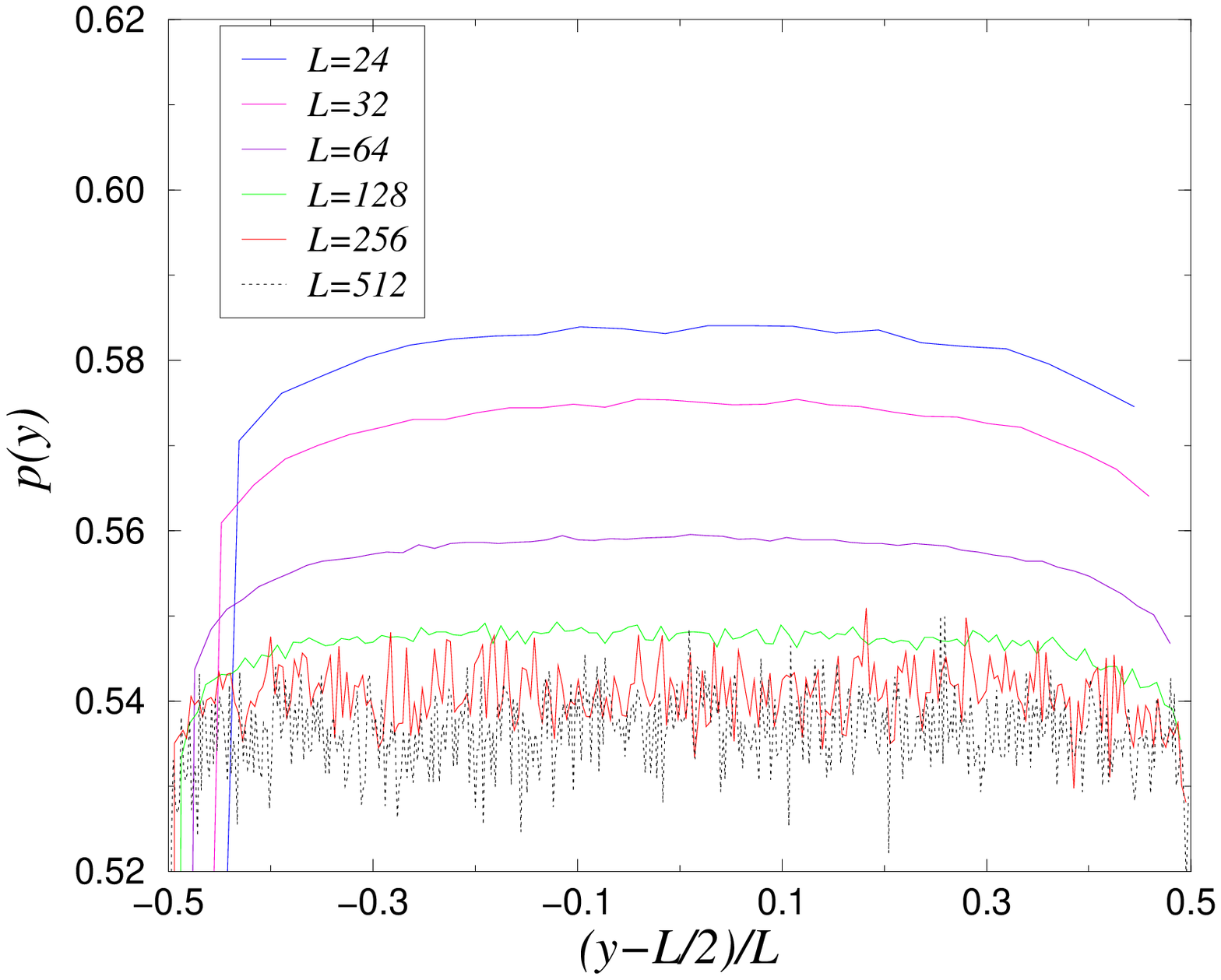,width=12cm,clip=!}}
\caption{Average damage profiles at peak load obtained by first centering the data
around the center of mass of the damage and then averging over different samples.
The data correspond to power law distributed disorder.}
\label{fig:loc3}
\end{figure}

We first consider the damage accumulated up to peak load shifting
the data by the center of mass method. The result displayed in
Fig.~\ref{fig:loc2} for the uniform disorder distribution clearly shows that there 
is no localization at peak load. In the case of power law disorder the
profile as presented in Fig.~\ref{fig:loc3} is not completely flat for small scales, but flattens more and
more as the size is increased. We tend thus to attribute the apparent
profile to size effects. These data imply that the localization profiles already observed
in the literature \cite{hansen003,hansen-other,mikko04} reflect 
mostly the final breakdown event, after the final spanning crack is 
nucleated. To quantify the corresponding damage localization, we can then 
average  the profiles $\Delta p(y)$  obtained considering only the damage accumulated 
between the peak load and failure. In addition, this procedure considerably reduces the 
background noise. 

\begin{figure}[hbtp]
\centerline{\psfig{file=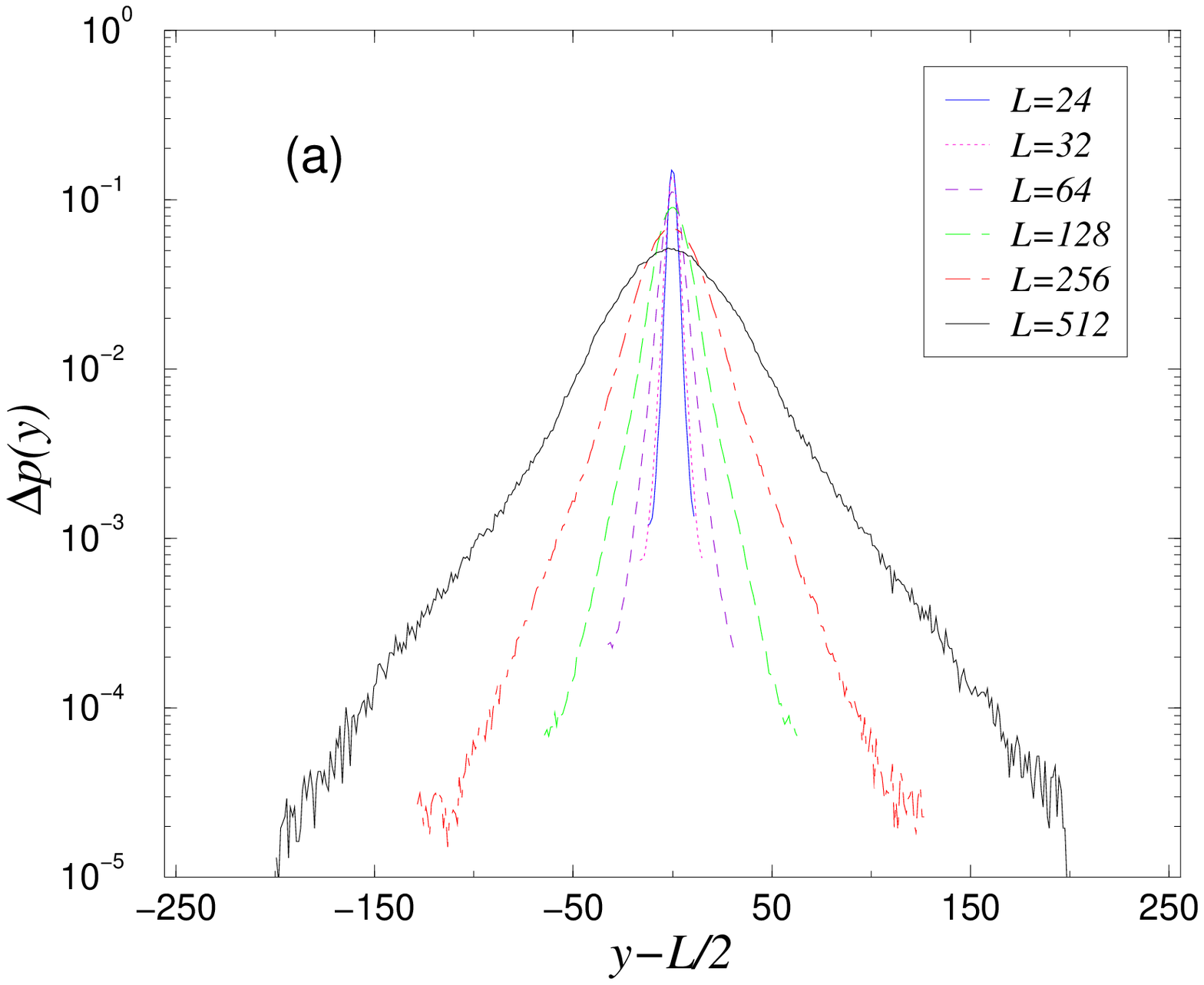,width=12cm,clip=!}}
\centerline{\psfig{file=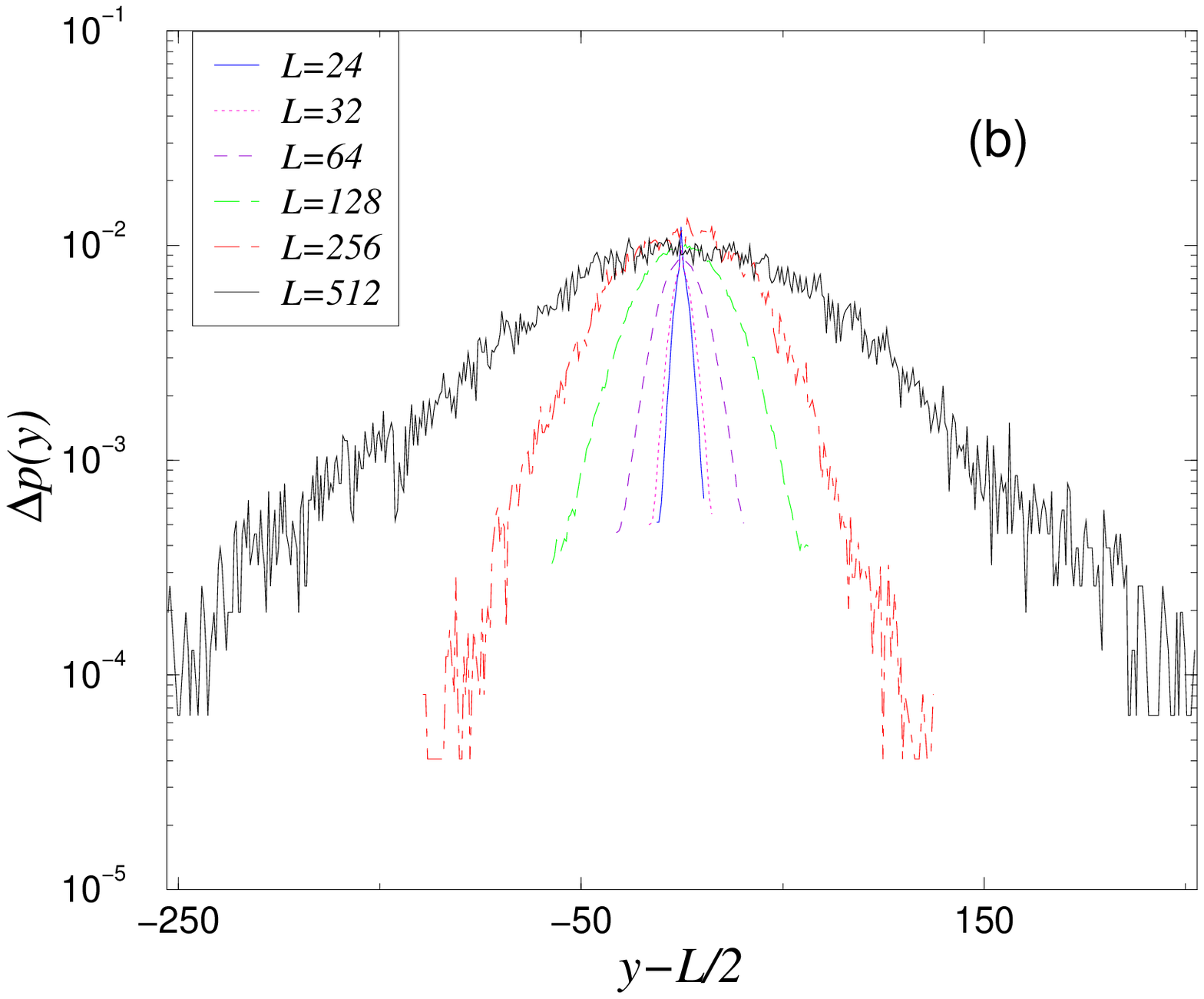,width=12cm,clip=!}}
\caption{Average profiles for the damage accumulated between peak load and failure. 
The average has been performed after shifting by the center of mass. The profiles
show exponential tails. 
(a) uniform disorder (b) power law disorder}
\label{fig:profcm}
\end{figure}

\begin{figure}[hbtp]
\centerline{\psfig{file=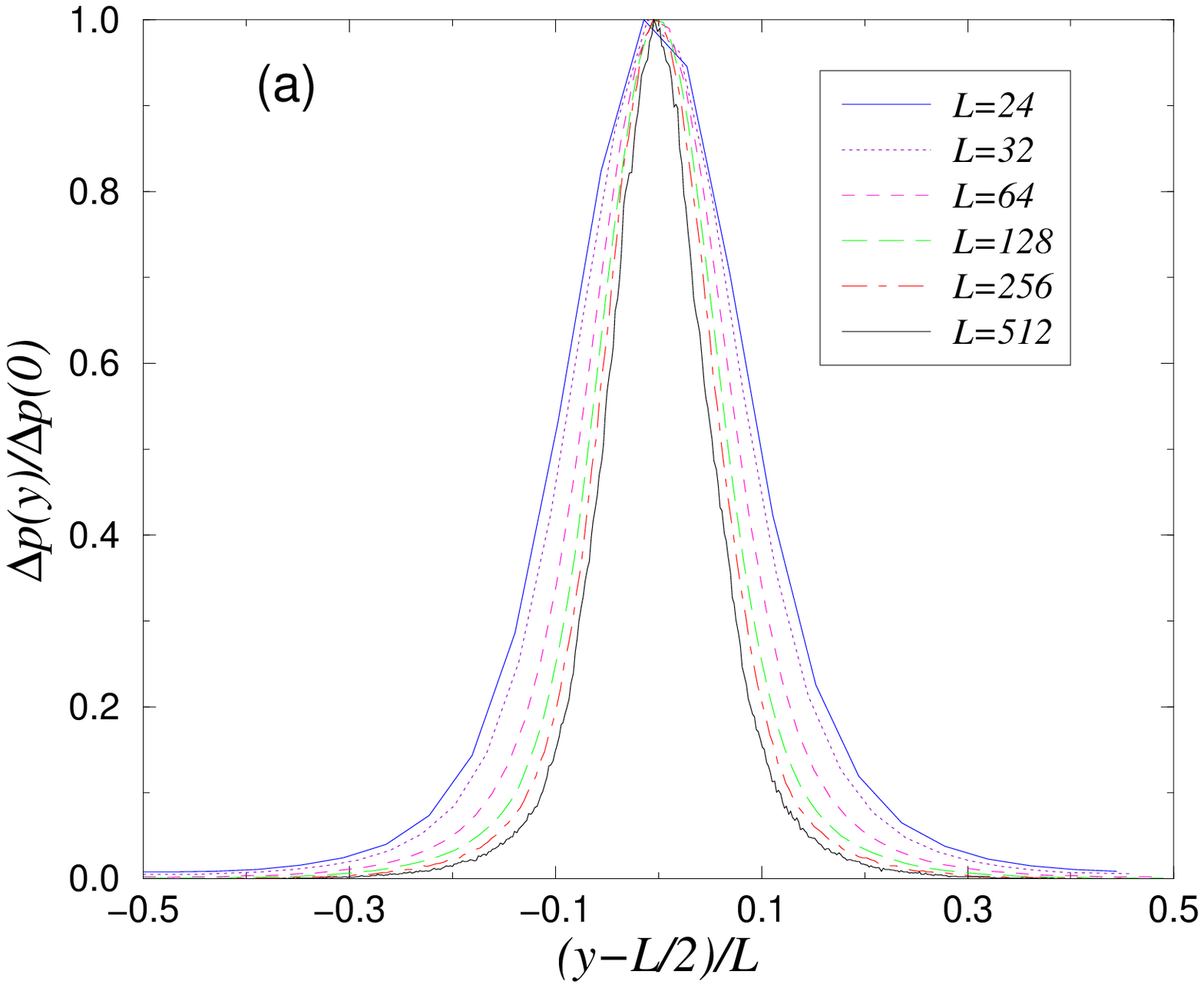,width=12cm,clip=!}}
\centerline{\psfig{file=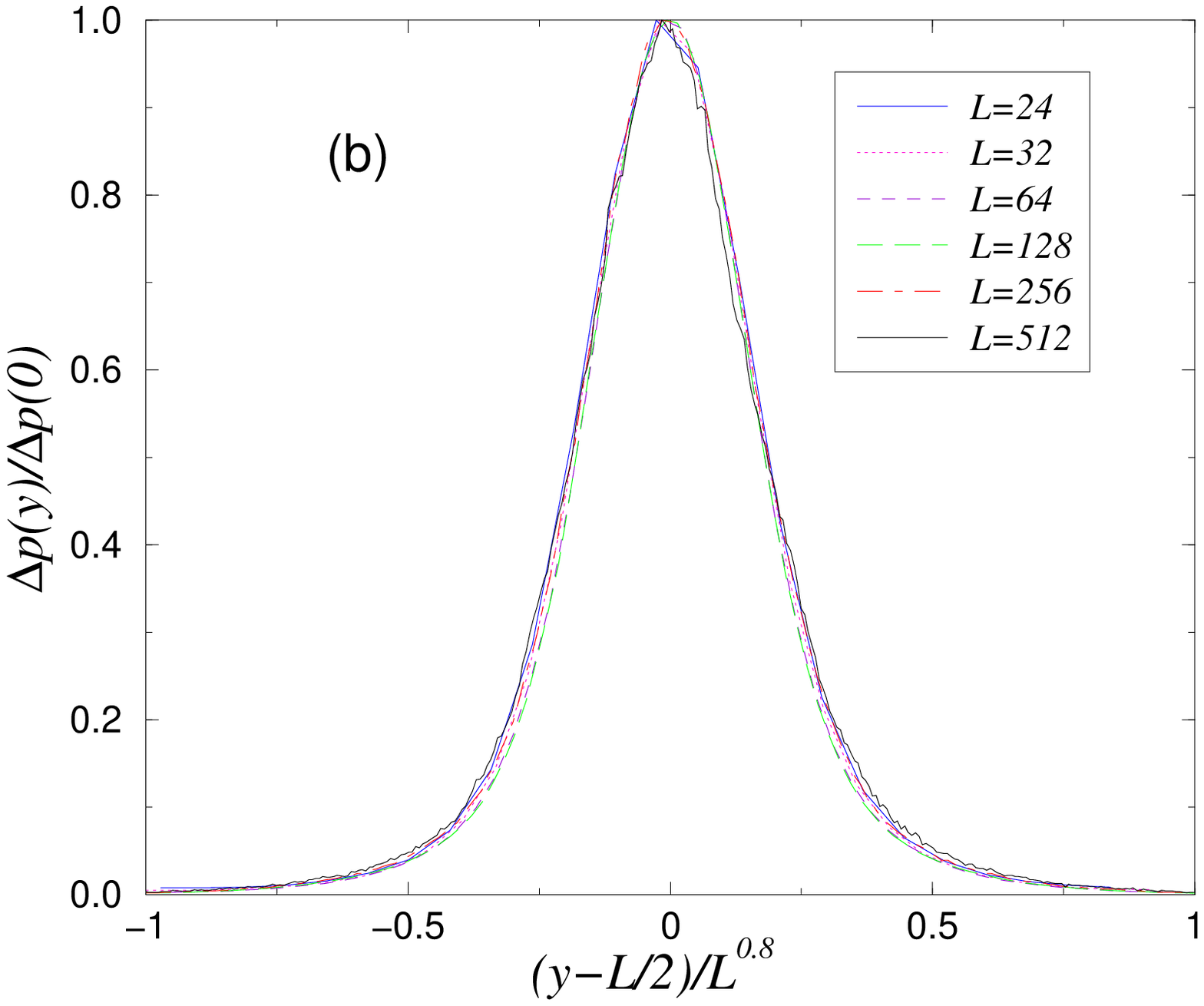,width=12cm,clip=!}}
\caption{Data collapse of the profiles reported in Fig.~\protect\ref{fig:profcm}a.
(a) Data collapse using a linear scaling for the localization
length. (b) Data collapse using a power law scaling.}
\label{fig:profCM_coll}
\end{figure}

The center of mass shifted averaged 
profiles for various system sizes are reported in Figs.~\ref{fig:profcm}a and 
~\ref{fig:profcm}b for uniform and power law disorder cases and these profiles 
show that the profile shapes decay exponentially at large system sizes. The 
damage peak $\langle \Delta p(0) \rangle$ scales as $L^{-0.3}$ for the uniform distribution
and is roughly constant for the power law distribution. 
We have tried different ways to collapse these profiles to 
extract a localization length. The simple linear scaling 
$\langle \Delta p(y,L) \rangle/\langle \Delta p(0) \rangle = f((y-L/2)/L)$, proposed in Ref.~ \cite{hansen003},
does not yield a very good collapse (see Fig.~\ref{fig:profCM_coll}a). 
A perfect collapse is instead obtained using the form
\begin{equation}
\langle \Delta p(y,L) \rangle/\langle \Delta p(0) \rangle= f(|y-L/2|/\xi),
\label{eq:prof2}
\end{equation}
where $\langle \Delta p(0) \rangle = L^{-0.3}$ and $\xi
\sim L^\alpha$, with $\alpha=0.8$ (see Fig.~\ref{fig:profCM_coll}b).
The situation for power law disored is similar, but scaling is less 
precise. As shown in Fig.~\ref{fig:profcm}b, the profiles are decaying
again exponentially. We could not, however, perform a reliable collapse:
a linear scaling seems appropriate for system sizes $L<256$ but fails
for larger sizes.

\begin{figure}[hbtp]
\centerline{\psfig{file=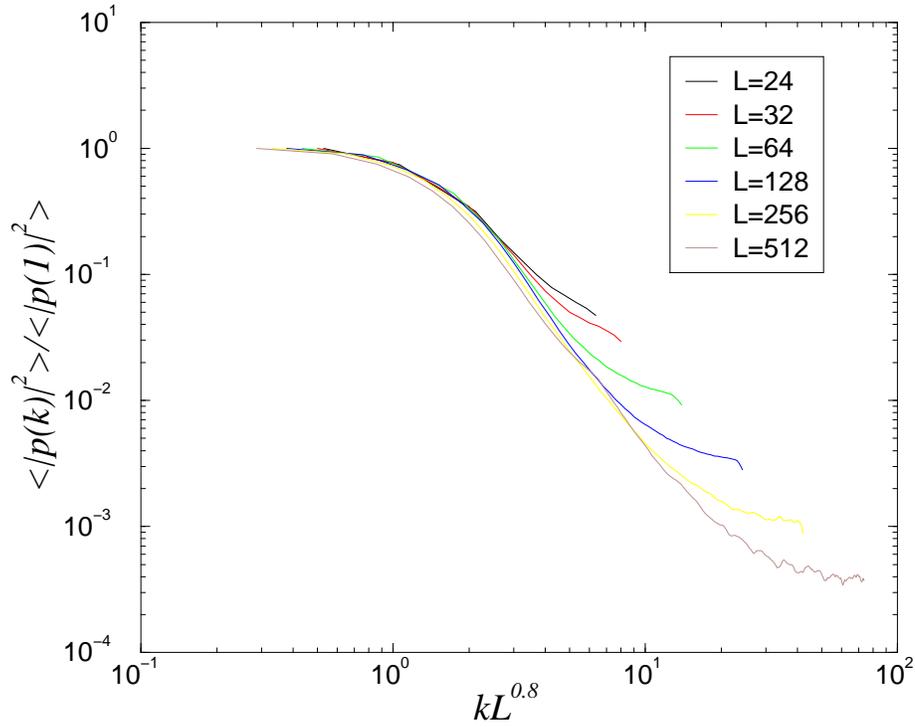,width=12cm,clip=!}}
\caption{Collapse of the power spectra of damage profiles for uniform disorder.}
\label{fig:loc7}
\end{figure}

To obtain additional confirmation of these results, we perform 
a Fourier analysis of the $\Delta p$ profiles, thus avoiding any possible 
bias due to the shifting. We first compute the magnitude of the Fourier transform for 
each realization and then average over 
disorder. From Eq.~\ref{eq:prof2} we would expect the
power spectrum of the profile to follow
\begin{equation}
\langle |\tilde{p}(k)|^2 \rangle/\langle |\tilde{p}(1)|^2 \rangle = \tilde{f}(k L^{0.8}).
\label{eq:prof2fft}
\end{equation}
This result applies to an infinite system and 
finite size deviations and other problems of the
discrete Fourier transform are expected to affect the
data. Nonetheless, as shown in Fig.~\ref{fig:loc7}, 
we can collapse reasonably well the curves using the same exponent 
as in real space. 

\begin{figure}[hbtp]
\centerline{\psfig{file=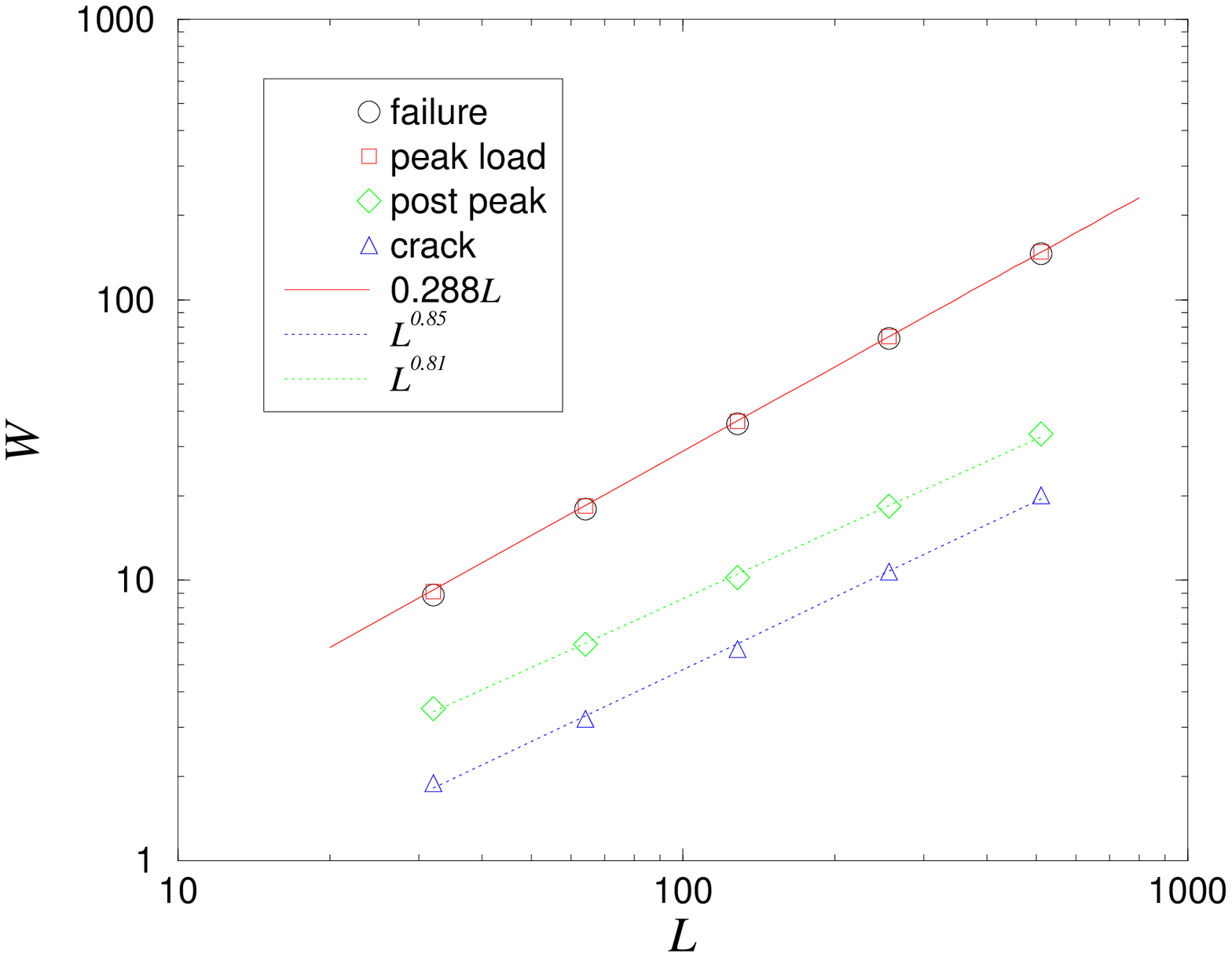,width=12cm,clip=!}}
\caption{The damage width at peak load and at failure are basically
the same. The linear scaling is expected for a uniform distribution
and it is not due to localization. On the other hand, localization
can be observed for post-peak damage, the width scales as a power law
similar to the one observed for the final crack.}
\label{fig:W}
\end{figure}

It is also possible to estimate the localization length directly,
independently of the profile averaging. This is done by computing the
width of the damage cloud as $W \equiv (\langle (y_b-\bar{y_b})^2
\rangle)^{1/2}$, where $y_b$ is the $y$ coordinate of a broken bond
and the average is taken over different realizations.  We have first
measured $W$ at peak load and at failure, and obtained a result 
$W \sim L$, consistent with earlier
results \cite{hansen91} (see Fig.~\ref{fig:W}).  This result is expected, since for a
uniformly distributed damage such as at the peak load, $W \simeq
L/\sqrt{12} \sim 0.288 L$, and is in excellent agreement with the
numerical data.  This is consistent with the fact that there is no
localization at peak-load and at failure the damage cloud is dominated
by the uniform fluctuations already present at peak load. Any
additional scaling due to localization is obscured. To uncover it we
can restrict the average only to the bonds broken in the last failure
event (post-peak damage), obtaining $W \sim L^{0.81}$ which is
consistent with the data collapse of the profiles. It is interesting
to notice that the width of the final crack scales as well with a similar exponent.
This supports the idea that localization is produced by catastrophic failure
which yields at the same time the crack and the damage profile.

\section{Scaling of the damage density}

\begin{figure}[hbtp]
\centerline{\psfig{file=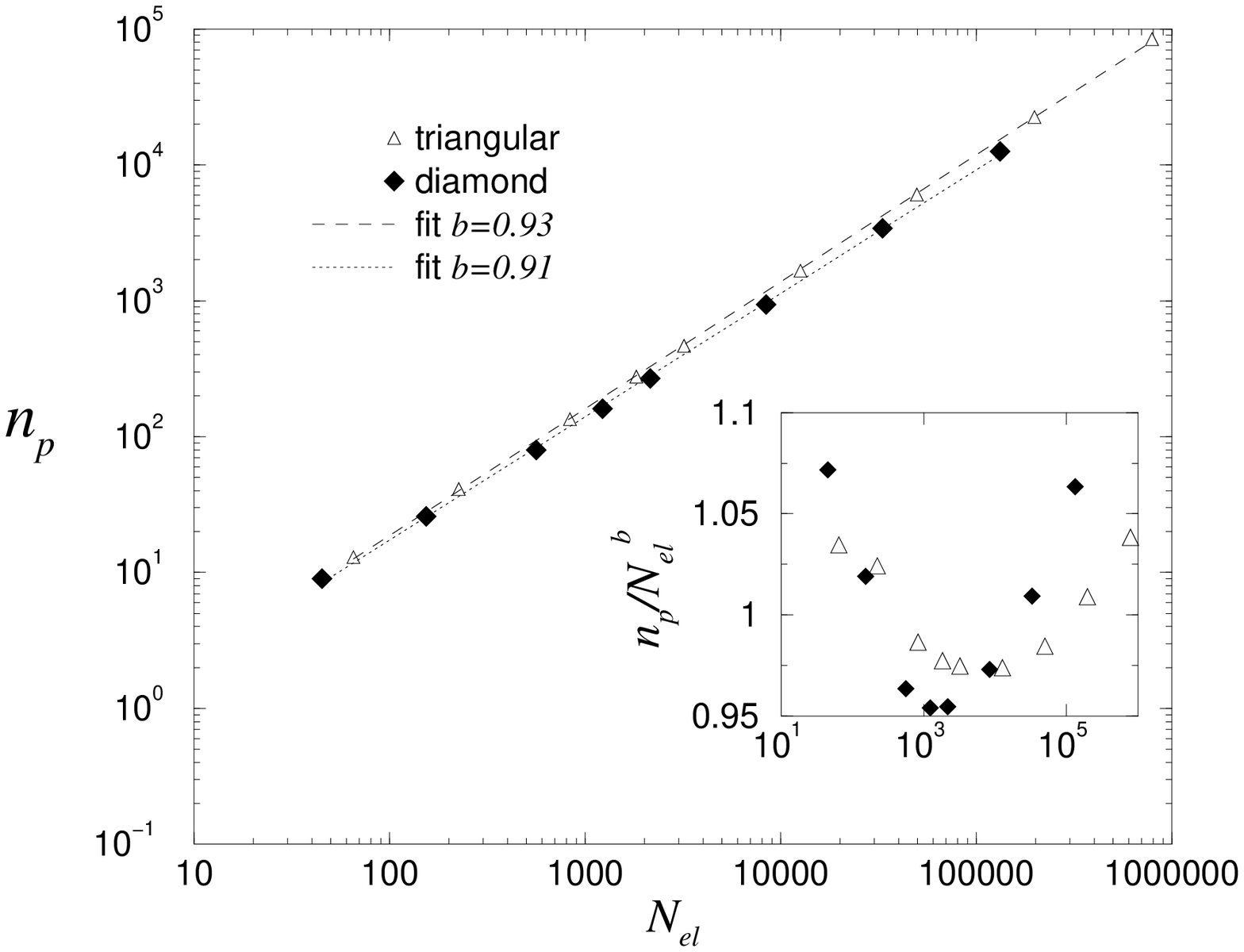,width=12cm,clip=!}}
\caption{Scaling of number of broken bonds at peak load for triangular
and diamond lattices. The scaling exponents are very close to each other
and the difference could be attributed to small size effects. There
are, apparently, some systematic errors as shown in the inset.}
\label{fig:2}
\end{figure}

\begin{figure}[hbtp]
\centerline{\psfig{file=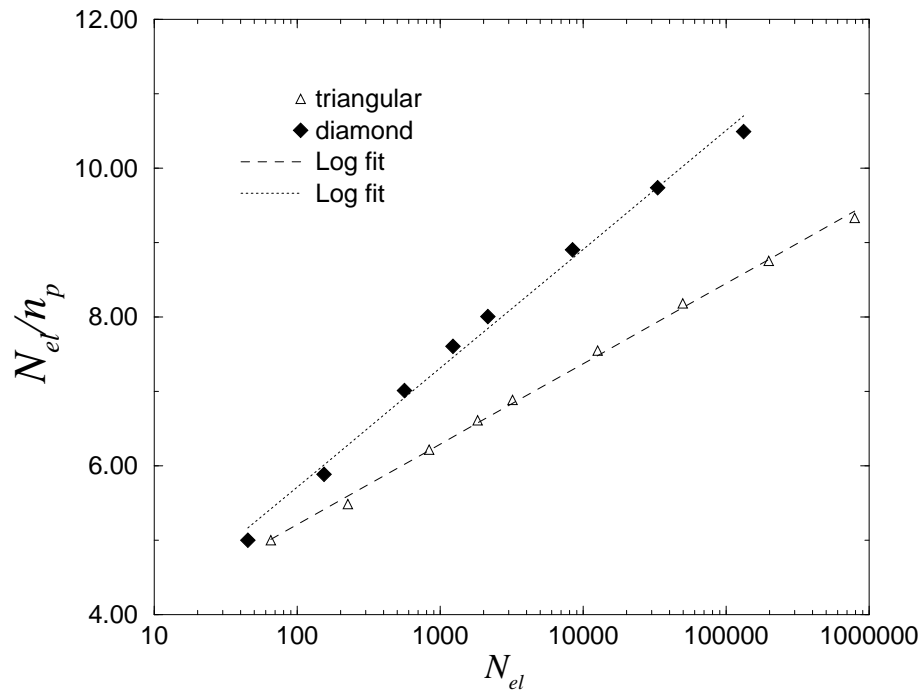,width=12cm,clip=!}}
\caption{The number of broken bonds at peak load can also be
fit by a linear function times a logarithmic correction by 
plotting $n_p/N_{el}$ as a function of $N_{el}$ in a log-linear plot.}
\label{fig:3}
\end{figure}

It has been noted in the previous section that the final breakdown
event is very different from the initial precursors. Thus, we consider
the scaling of the number of broken bonds at the peak load, $n_p$,
that excludes the last catastrophic event. In Fig.~\ref{fig:2} we plot
$n_p$ as a function of the lattice size $N_{el}$ for triangular and
diamond lattices. The data displays a reasonable power law behavior
$n_p \sim N_{el}^b$, with $b=0.93$ and $b=0.91$ for triangular and
diamond lattice, respectively as previously shown in
Ref.~\cite{delaplace}. Thus the difference between the two lattices is
marginal and may be attributed to the results obtained from the
smaller lattice sizes, where corrections to the fractal scaling may
exist.  By plotting $n_p/N_{el}^b$ vs $N_{el}$ (see the inset of
Fig.~\ref{fig:2}) we show that some systematic deviations appear.
The data could be equally well fit by a linear law times a
logarithmic correction $n_p \simeq N_{el}/\log(N_{el})$ as suggested
in Ref.~\cite{mikko04} (see Fig.~\ref{fig:3}).  Both these fits imply
that in the limit of large lattices the fraction of broken bonds prior
to fracture vanishes (i.e. $p_c=0$ in the thermodynamic limit).

\begin{figure}[hbtp]
\centerline{\psfig{file=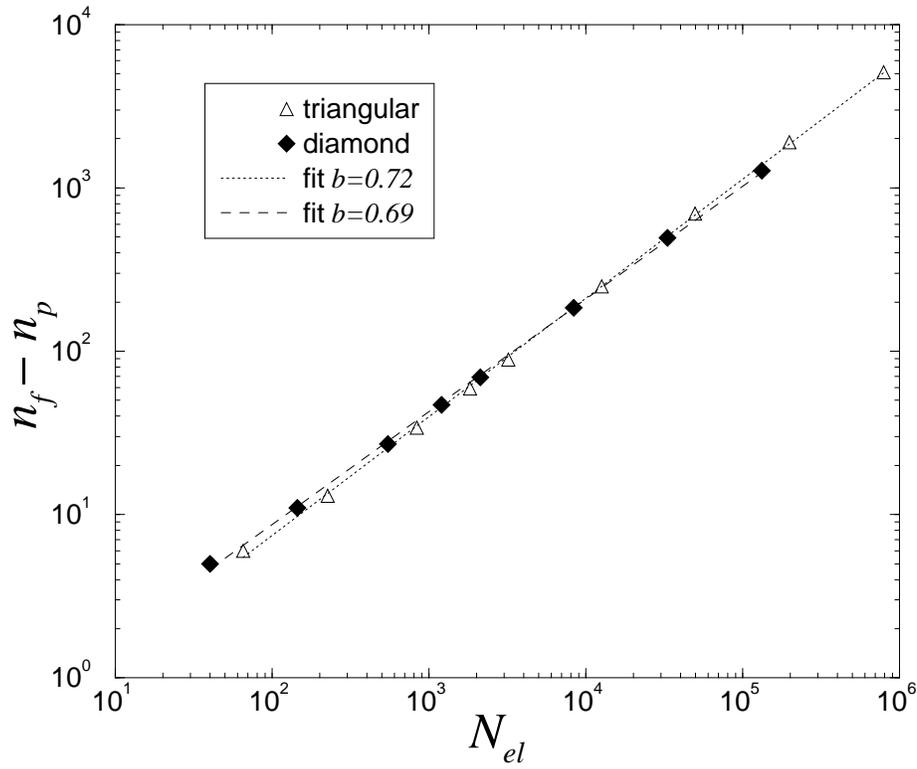,width=12cm,clip=!}}
\caption{The number of bonds broken in the last catastrophic event
scales as a power law of $N_{el}$.}
\label{fig:4}
\end{figure}

\begin{figure}[hbtp]
\centerline{\psfig{file=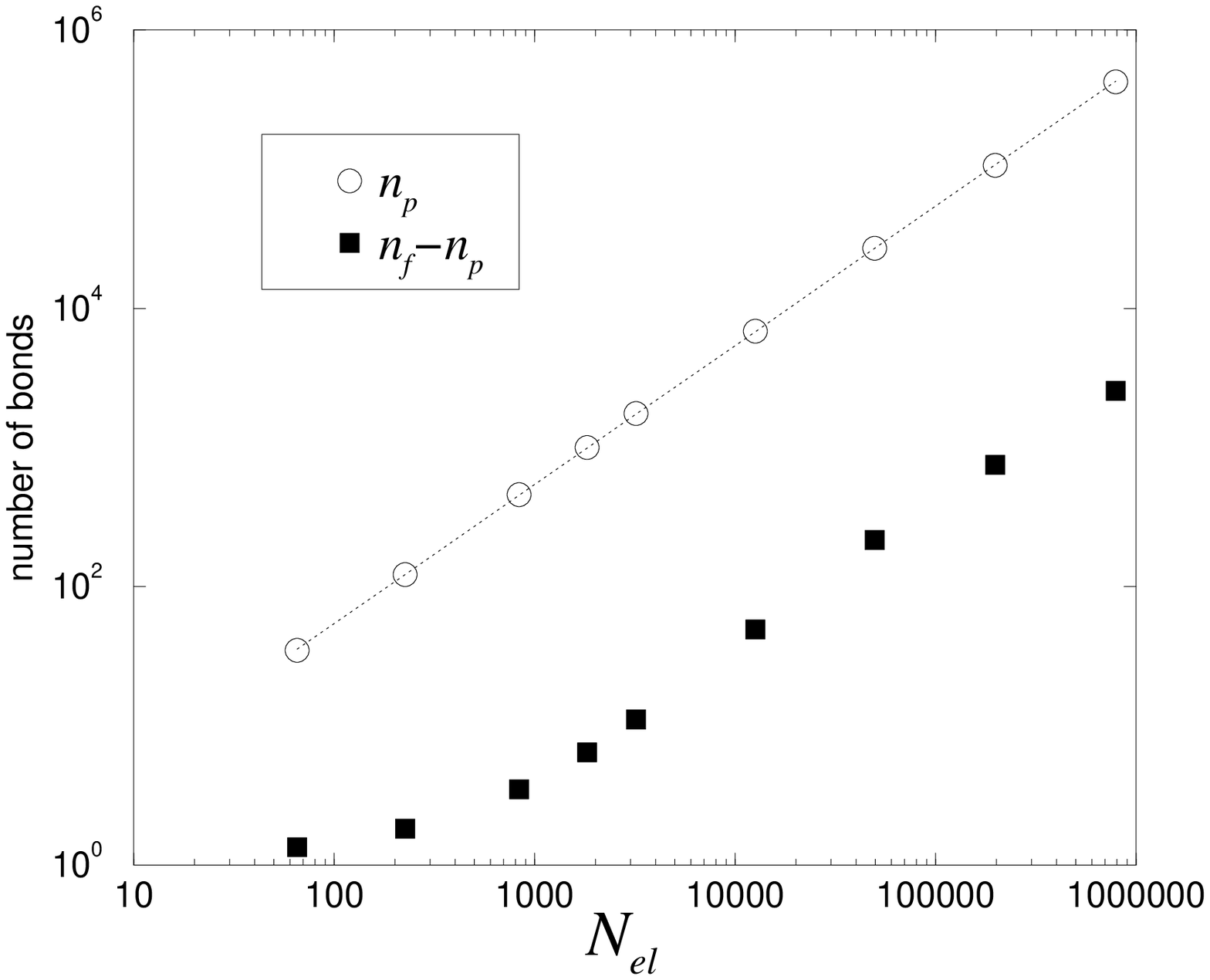,width=12cm,clip=!}}
\caption{Scaling of number of broken bonds for a triangular lattice
with power law threshold distribution. The scaling exponents for $n_p$
is very close to one and a similar result seems to hold asymptotically
for $n_f-n_p$.}
\label{fig:4a}
\end{figure}

Figure \ref{fig:4} presents the scaling for the number of broken
bonds, $(n_f-n_p)$, after crossing the peak load. Once again, the
scaling exponent for $(n_f-n_p)$ is similar for both the triangular
and diamond lattice topologies, and is equal to $0.72$ and $0.69$
respectively.  This exponent is consistent with the parameters
estimated from the profile $(n_f-n_p) \sim <p(0)>(3L+1)\xi \sim L^{1.5}$, 
The behavior of power law disorder appears to be
simpler. The number of broken bonds at peak load scales linearly with
$N_{el}$, as shown in Fig.~\ref{fig:4a}. A similar result holds
asymptotically for $(n_f-n_p)$ (see Fig.~\ref{fig:4a}), which is
again consistent with the profile scaling.


\section{The failure probability distribution}

\begin{figure}[hbtp]
\centerline{\psfig{file=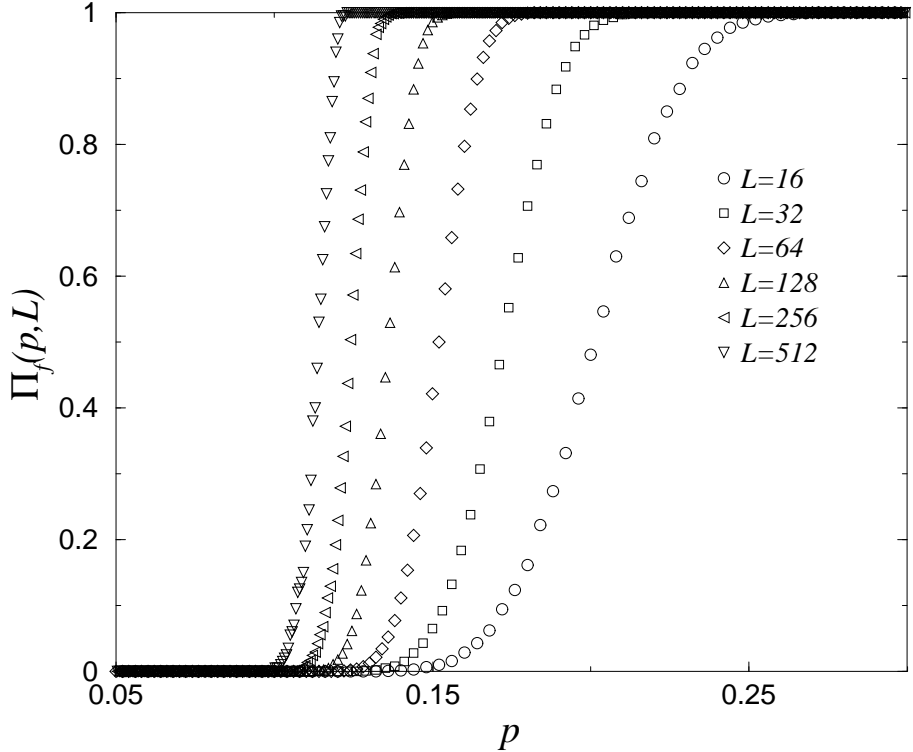,width=12cm,clip=!}}
\caption{The cumulative probability distribution for the fraction of broken
bonds at failure for triangular lattices of different system sizes.}
\label{fig:5}
\end{figure}

The cumulative probability distribution for the fraction of broken
bonds at failure (also termed as cumulative failure probability
distribution), defined as the probability $\Pi_{f} (p_b,L)$ that a
system of size $L$ fails when the fraction of broken bonds equals $p_b
= \frac{n_b}{N_{el}}$, where $n_b$ is the number of broken bonds, is
plotted in Fig. ~\ref{fig:5} for varying triangular lattice system
sizes. In Ref.~\cite{hansen003} a data collapse of a different, but,
related distribution (i.e., survival probability) was attempted using
percolation scaling. As it is evident from the failure of such a
scaling (i.e., Eq. (\ref{pf})), this collapse is poor.

\begin{figure}[hbtp]
\centerline{\psfig{file=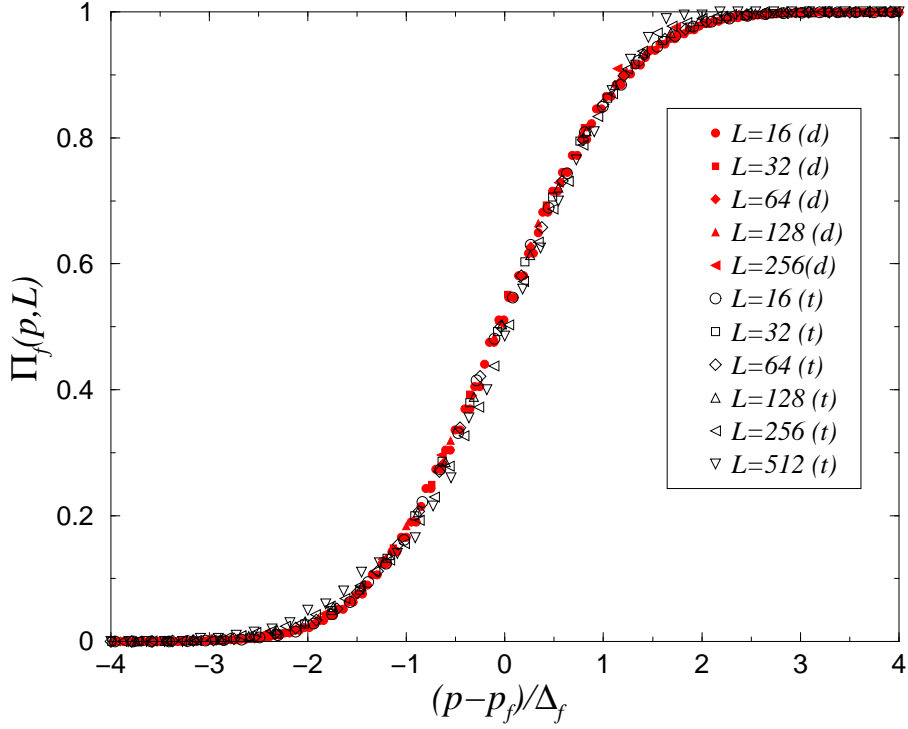,width=12cm,clip=!}}
\caption{The collapsed cumulative failure probability distributions 
for both triangular (t) and diamond (d) lattices of different system sizes
with uniform disorder  when plotted as a function of the reduced variable 
$\bar{p}_f=(p-p_f)/\Delta_f$.}
\label{fig6}
\end{figure}

\begin{figure}[hbtp]
\centerline{\psfig{file=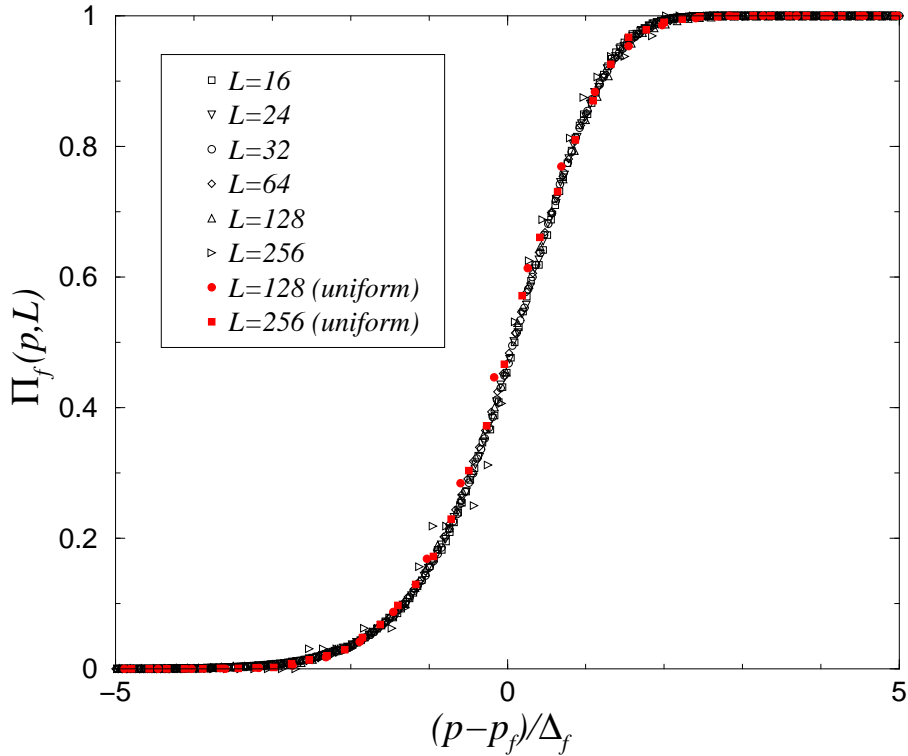,width=12cm,clip=!}}
\caption{The collapsed cumulative failure probability distributions
for lattices of different system sizes with power law disorder when
plotted as a function of the reduced variable
$\bar{p}_f=(p-p_f)/\Delta_f$. For comparison we also include
two curves obtained in the case of uniform disorder.}
\label{fig7}
\end{figure}

\begin{figure}[hbtp]
\centerline{\psfig{file=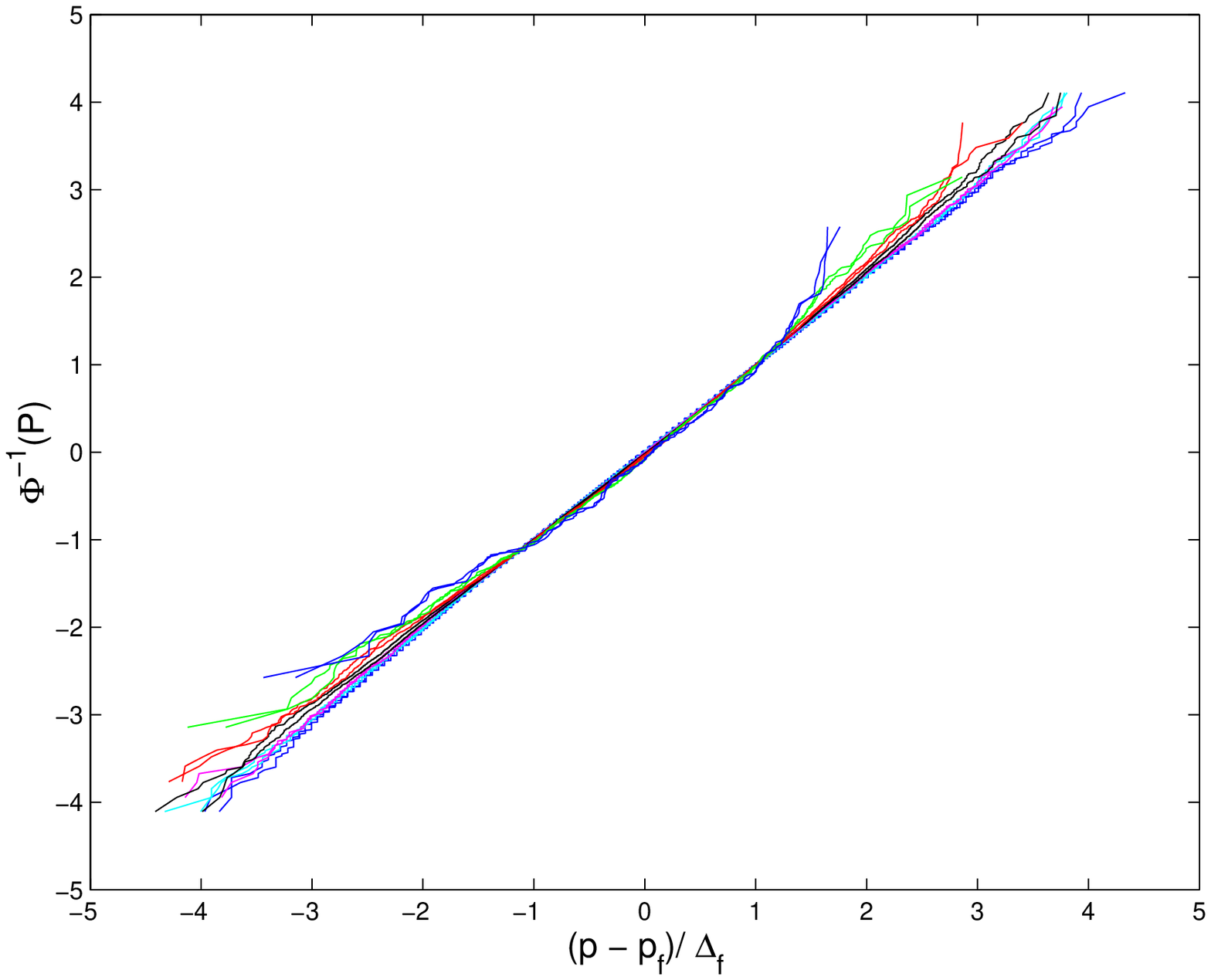,width=12cm,clip=!}}
\caption{Normal distribution fit for the cumulative probability distributions 
of the fraction of broken bonds at failure and at the peak load 
for triangular lattices of different system sizes $L$ = \{16, 24, 32, 64, 128, 256, 512\}.}
\label{fig9}
\end{figure}

On the other hand, we obtain a very good collapse by simply plotting
the distribution in terms of $\bar{p}_f \equiv \frac{(n_b -
\mu_{n_f})}{\sigma_{n_f}} = \frac{(p_b - p_{f})}{\Delta_{p_f}}$, where
$\mu_{n_f}$ and $\sigma_{n_f}$ denote the mean and standard deviation
of the number of broken bonds at failure, and $p_f$ and $\Delta_{p_f}$
denote the mean and standard deviation of fraction of broken bonds at
failure.  Fig.~\ref{fig6} shows that $\Pi_{f} (p,L)$ may be
expressed in a universal scaling form such that $\Pi_{f} (p,L) =
\Pi_{f} (\bar{p}_f)$ for both triangular and diamond lattice
topologies of different system sizes $L$.  A similar collapse
can be performed for the power law disorder distribution (see
Fig.~\ref{fig7}). The excellent collapse of
the data in Figs. \ref{fig6} and \ref{fig7}
 suggests that the cumulative failure
probability distribution, $\Pi_{f} (p_b,L) = \Pi_{f} (\bar{p}_f)$, may
be universal in the sense that it is independent of lattice topology
and disorder distribution.We have also checked that the distributions at peak load obey
essentially the same laws, i.e., $\Pi (\bar{p}) = \Pi_{f} (\bar{p}_f)
= \Pi_{p} (\bar{p}_p)$, where $\Pi_{p} (\bar{p}_p)$ is the probability
that a system of size $L$ is at the peak load when the fraction of
broken bonds equals $p_b$, and $\bar{p}_p$ is the corresponding
reduced variable at the peak load.  Finally, the collapse of the data
in Fig. \ref{fig9} indicates that a Gaussian distribution adequately
describes $\Pi_f$.

The fact that damage is Gaussian distributed suggests that there is no
divergent correlation length at failure consistent with the
conclusions of Ref.~\cite{delaplace} that reported a finite
correlation length at failure. Long-range correlations in the damage
would imply that the central limit theorem does not hold and hence the
normal distribution would not be an adequate fit to the data. The
absence of long-range corralation is again in agreement with the
hypothesis that fracture is analogous to a first-order transition
\cite{zrvs}.

\section{Discussion}

This paper presents numerical simulations on large two-dimensional
triangular and diamond lattice fuse networks with uniform and
power law disorder distributions. We focus our analysis on
the statistical properties and localization features of the
damage as a function of the lattice type, size and disorder
form. The use of high statistical sampling and relatively large
lattice sizes is essential to obtain reliable results.

The picture emerging from our analysis is that, for strong 
disorder, damage accumulates first in the system in an uncorrelated
(or short range correlated) manner. This process continues up to the 
peak load, where no apparent sign of localization is present.
Further increase of the current leads to catastrophic failure
through a large avalanche event, whose size scales with the lattice size
as $L^{1.45}$.  The damage accumulated in this event is localized
in a cloud surrounding the final crack. The damage profile has
exponential tails and can be collapsed for different system sizes
using an appropriate scaling law.

The accumulated damage density at failure $p_f$ appears to be far from
a percolation like critical point, since we could not find a reliable
universal scaling law as the system size is varied. A different 
possibility that we have tested is that $p_f\to 0$ as $L\to \infty$,
but a fit using simple scaling forms show some deviation as well.
Thus we can not conclude whether $p_f$ decays slowly to zero, or to
a non vanishing asymptotic value. This value, however, would not 
necessarily coincide with a critical point. 

Finally, this study also presents the scaling of
cumulative failure probability distribution, which is defined as the
probability that a lattice system fails at a given fraction of broken
bonds. Based on the numerical results presented, we show that the
cumulative failure probability distribution is universal in the sense
that it does not depend on the lattice topology, i.e., the
distributions are identical for both triangular and diamond lattice
topologies. Furthermore, a {\it normal} distribution presents an
adequate fit to the data. 
 
\ack 
%\section*{Acknowledgment}
This research is sponsored by the
Mathematical, Information and Computational Sciences Division, Office
of Advanced Scientific Computing Research, U.S. Department of Energy
under contract number DE-AC05-00OR22725 with UT-Battelle, LLC. We
thank M. J. Alava, A. Hansen, H. J. Herrmann and S. Roux for useful remarks and
discussions.

\References
%\begin{thebibliography}{99}

\bibitem{breakdown} Herrmann HJ and Roux S (eds.), {\em
        Statistical Models for the Fracture of Disordered Media},
        (North-Holland, Amsterdam, 1990).   Bardhan KK,
         Chakrabarti BK and  Hansen A (eds.), {\em Non-linearity and
        breakdown in soft condensed matter}, (Springer Verlag, Berlin,
        1994).  Chakrabarti  BK and Benguigui LG, {\em
        Statistical physics of fracture and breakdown in disordered
        systems} (Oxford Univ.  Press, Oxford, 1997).   Krajcinovic D
        and van Mier, {\em Damage and fracture of disordered
        materials}, (Springer Verlag, New York, 2000).

\bibitem{man}
        Mandelbrot BB, Passoja DE, and  Paullay AJ 1984
        Nature (London) {\bf 308}, 721 
\bibitem{bouch}
        For a review see Bouchaud E 1997 J Phys. C {\bf 9}, 4319 

\bibitem{ciliberto}
        Garcimart\'{\i}n A,  Guarino A,
         Bellon L and Ciliberto S 1997
        Phys. Rev. Lett. {\bf 79}, 3202 ;
       Guarino A , Garcimart\'{\i}n  A and  Ciliberto S 1998
        Eur. Phys. J. B {\bf 6}, 13 .
\bibitem{strauven}
         Maes C,  Van Moffaert A, Frederix H and  Strauven H 1998
        Phys. Rev. B {\bf 57}, 4987.
\bibitem{ae}
        Petri A, Paparo G,  Vespignani A 1994
        Alippi A and Costantini M,
        Phys. Rev. Lett. {\bf 73}, 3423 
\bibitem{paper}
        Salminen LI, Tolvanen AI , and Alava MJ 2002
        Phys. Rev. Lett. {\bf 89}, 185503 
\bibitem{hansen91b}
Hansen A, Hinrichsen EL, and Roux S, 1991
Phys. Rev. Lett. {\bf 66}, 2476 

\bibitem{bat-98}
Batrouni GG and Hansen A 1998
	Phys. Rev. Lett. {\bf 80}, 325 (1998).

\bibitem{rai-98}
  R\"ais\"anen VI, Sepp\"al\"a ET, Alava MJ 1988
  and Duxbury PM, Phys. Rev. Lett. 80, 329 (1998).
\bibitem{sep-00}
Sepp\"al\"a ET, R\"ais\"anen VI, and  Alava MJ 2000
Phys. Rev. E {\bf 61}, 6312 (2000)

\bibitem{hansen}
        Hansen A and Hemmer PC 1994
        Phys. Lett. A {\bf 184}, 394 (1994).
\bibitem{zrvs}
        Zapperi S, Ray P, Stanley HE, and Vespignani A 1997
        Phys. Rev. Lett. {\bf 78}, 1408; 1999
        Phys.\ Rev.\ E {\bf 59}, 5049

\bibitem{zvs}
        Zapperi S, Vespignani A, and Stanley HE 1997
        Nature (London) {\bf 388}, 658 
\bibitem{gcalda}
        Caldarelli G, Di Tolla FG and Petri A 1996
        Phys. Rev. Lett. {\bf 77}, 2503 
\bibitem{alava}
         R\"ais\"anen VI, Alava MJ, Nieminen RM 1998
         Phys. Rev. B, {\bf 58}, 14288 
\bibitem{kahng88}
        Kahng B, Batrouni GG, Redner S, de Arcangelis L
        and Herrmann HJ 1988 Phys. Rev. B {\bf 37}, 7625
\bibitem{delaplace}
	Delaplace A, Pijaudier-Cabot G, and Roux S 1996
	Journal of Mechanics and Physics of Solids, {\bf 44}, 99 
\bibitem{deArcangelis89}
	de Arcangelis L, Hansen A, Herrmann HJ, and Roux S 1989
	Phys. Rev. B, {\bf 40}, 877 
\bibitem{nukalajpamg}
	Nukala PKVV, and Simunovic S 2003
	J. Phys. A: Math. Gen. {\bf 36}, 11403 
\bibitem{deArcangelis85}
	de Arcangelis L, Redner S, and Herrmann HJ 1985
	J. Phys. (Paris) Lett. {\bf 46} 585; 
	Sahimi M and Goddard JD 1986
	Phys. Rev. B, {\bf 33}, 7848.
\bibitem{sornette}
	Andersen JV, Sornette D, and Leung KT 1997
	Phys. Rev. Lett. {\bf 78}, 2140 
	Sornette D and Andersen JV 1998
	Euro. Phys. Journal B. {\bf 1} 353 
	Johansen  A and Sornette D 2000
	Euro. Phys. Journal B. {\bf 18} 163 

\bibitem{guyon88}
	Roux S, Hansen A, Herrmann HJ, and Guyon E 1988
	J. Stat. Phys. {\bf 52}, 237 

\bibitem{hansen003}
	Hansen A and Schmittbuhl J 2003
	Phys. Rev. Lett. {\bf 90}, 45504 
\bibitem{hansen-other}
  Bakke JOH, Bjelland J, Ramstad T, Stranden T
  Hansen A and Schmittbuhl 2003 Physica Scripta {\bf T106} 65

\bibitem{mikko04}
  Reurings F and Alava MJ 2004
  preprint cond-mat/0401592

\bibitem{ramstad}
Ramstad T et al. , preprint cond-mat/0311606

\bibitem{batrouni}
	Batrouni GG, Hansen A, and Nelkin M 1986 
	Phys. Rev. Lett. {\bf 57}, 1336; 1988
	J. Stat. Phys. {\bf 52}, 747 
	
\bibitem{hansen91}
	Hansen A, Hinrichsen EL and Roux S 1991 
	Phys. Rev. B, {\bf 43}, 665 
\bibitem{hansen001}
	Hansen  A and Roux S 2000, 
	{\em Statistical toolbox for damage and fracture}, 17-101, in book 
	{\em Damage and Fracture of Disordered Materials}, eds.  
	D. Krajcinovic and van Mier, Springer Verlag, New York.

%\endbib
%\end{thebibliography}
\endrefs

\begin{table}[hbtp]
  \leavevmode
  \begin{center}
  \vspace*{1ex}
  \begin{tabular}{|c|c|c|c|c|c|c|c|c|c|}\hline
  L  & $N_{config}$ & \multicolumn{4}{c|}{Triangular} & \multicolumn{4}{c|}{Diamond}\\\cline{3-10}
     &  & $p_{p}$ & $\Delta_{p}$ & $p_{f}$ & $\Delta_{f}$ & $p_{p}$ & $\Delta_{p}$ & $p_{f}$ & $\Delta_{f}$ \\
  \hline
  4 & 50000  & 0.2070 & 0.0532 & 0.3030 & 0.0476 & 0.2367 & 0.0625 & 0.3611 & 0.0482 \\
  8 & 50000  & 0.1813 & 0.0346 & 0.2440 & 0.0329 & 0.1794 & 0.0365 & 0.2576 & 0.0318 \\
 16 & 50000  & 0.1612 & 0.0225 & 0.2023 & 0.0218 & 0.1470 & 0.0222 & 0.1972 & 0.0202 \\
 24 & 50000  & 0.1513 & 0.0177 & 0.1841 & 0.0170 & 0.1340 & 0.0169 & 0.1731 & 0.0155 \\
 32 & 50000  & 0.1451 & 0.0150 & 0.1731 & 0.0143 & 0.1267 & 0.0139 & 0.1596 & 0.0129 \\
 64 & 50000  & 0.1325 & 0.0104 & 0.1524 & 0.0096 & 0.1132 & 0.0092 & 0.1353 & 0.0084 \\
128 & 12000  & 0.1222 & 0.0078 & 0.1362 & 0.0070 & 0.1031 & 0.0064 & 0.1181 & 0.0056 \\
256 & 1200  & 0.1142 & 0.0058 & 0.1238 & 0.0053 & 0.0955 & 0.0048 & 0.1052 & 0.0042 \\
512 & 200  & 0.1072 & 0.0048 & 0.1136 & 0.0044 &  &  \\
  \hline
  \end{tabular}
  \label{table2}
  \end{center}\caption{A summary of the main results of the simulations for uniform thresholds 
distribution, including the number of configurations used to average the results for each system size. 
$p_{p}$ and $p_{f}$ denote the mean fraction of broken bonds in a lattice system of size $L$ 
at the peak load and at failure, respectively. Similarly, $\Delta_{p}$ and $\Delta_{f}$ denote the 
standard deviation of fraction of broken bonds at the peak load and at failure respectively.}

\end{table}

\begin{table}[hbtp]
  \leavevmode
  \begin{center}
  \vspace*{1ex}
  \begin{tabular}{|c|c|c|c|c|c|}\hline
  L  & $N_{config}$ & \multicolumn{4}{c|}{Triangular} \\\cline{3-6}
     &  & $p_{p}$ & $\Delta_{p}$ & $p_{f}$ & $\Delta_{f}$ \\
  \hline
  4 & 50000  & 0.5360 & 0.0450 & 0.5568 & 0.0437 \\
  8 & 50000  & 0.5454 & 0.0324 & 0.5535 & 0.0318 \\
 16 & 50000  & 0.5489 & 0.0217 & 0.5531 & 0.0213 \\
 24 & 50000  & 0.5491 & 0.0170 & 0.5526 & 0.0166 \\
 32 & 50000  & 0.5483 & 0.0143 & 0.5517 & 0.0139 \\
 64 & 25000  & 0.5449 & 0.0096 & 0.5489 & 0.0092 \\
128 & 1400  & 0.5406 & 0.0082 & 0.5449 & 0.0080 \\
256 & 32  & 0.5379 & 0.0045 & 0.5417 & 0.0040 \\
512 & 10  & 0.5349 & 0.0037 & 0.5382 & 0.0032 \\
  \hline
  \end{tabular}
  \label{tableD20}
  \end{center}\caption{A summary of the main results of the simulations for the power law thresholds 
distribution, including the number of configurations used to average
the results for each system size.  $p_{p}$ and $p_{f}$ denote the mean
fraction of broken bonds in a lattice system of size $L$ at the peak
load and at failure, respectively. Similarly, $\Delta_{p}$ and
$\Delta_{f}$ denote the standard deviation of fraction of broken bonds
at the peak load and at failure respectively.}

\end{table}

\end{document}